\documentclass[preprint,12pt]{elsarticle}

\usepackage{packages}
\journal{}

\begin{document}

\begin{frontmatter}



\title{IBPA: Real-time Free-form Manifold Mesh Reconstruction via Incremental Ball Pivoting with Integrated Hole Detection}


\author[inst1]{Mauhing Yip}

\affiliation[inst1]{organization={Department of Engineering Cybernetics, NTNU},%Department and Organization
            addressline={O. S. Bragstads Plass 2D}, 
            city={Trondheim},
            postcode={7034}, 
            country={Norway}}

\author[inst1]{Mohit Singh}
\author[inst1]{Kostas Alexis}
\author[inst2]{Christian Schellewald}
\author[inst1]{Annette Stahl}

\affiliation[inst2]{organization={SINTEF Ocean},%Department and Organization
            addressline={Brattørkaia 17c}, 
            city={Trondheim},
            postcode={7010}, ,
            country={Norway}}

\begin{abstract}
Both Remotely Operated underwater Vehicles (ROVs) and Autonomous Underwater Vehicles (AUVs) are frequently deployed to acquire geometric bathymetric data. However, it is often discovered post-survey that the acquired data coverage is incomplete. Given the high operational cost associated with underwater deployments, it is essential to incrementally visualize surface coverage in real-time to support informed decision-making by both the operators of ROVs  and the AUVs during data collection. In addition, traditional incremental surface reconstruction methods, such as Digital Terrain Models (DTMs), are inherently limited in expressiveness: they represent surfaces as height fields, allows only one elevation value per $(x, y)$ coordinate and thus cannot capture overhangs or vertical structures.

To overcome these limitations, we adapt the original Ball Pivoting Algorithm (BPA) into an incremental, real-time, and free-form surface reconstruction method, referred to as Incremental BPA (IBPA). Our method incrementally constructs an orientable, manifold mesh from streaming point cloud data without imposing assumptions regarding point cloud overlap or spatial distribution.
Furthermore, we introduce a hole detection mechanism that identifies and highlights incomplete mesh regions. Compared to existing approaches, our method supports more complex surface topologies without prior structural assumptions.

The source code of our reference implementation is available: \url{https://github.com/Mauhing/Incremental-BPA}
\end{abstract}
%\centering
%\includegraphics[scale=0.7]{grabs.png}

%    \item Our surface reconstruction method generates meshes that are orientable and manifold, including both vertex-manifold and edge-manifold properties.
%    \item The method supports free-form surface reconstruction and can represent overhangs, vertical walls, and complex geometries where multiple $z$ values exist for the same $(x, y)$.
%    \item It is versatile and compatible with various 3D point cloud acquisition methods, including those with variable resolution.
%    \item The method is robust against outliers in the point cloud.
%    \item It is applicable to both indoor and outdoor environments.
%    \item The algorithm does not fabricate or hallucinate points to improve visual appearance of the surface.
%    \item Our online reconstruction method supports hole detection, providing feedback to operators of remote operated vehicle or autonomous underwater vehicles (AUVs) about areas with insufficient geometric data in real-time—before the mission is finished.
%    \item The method makes no assumptions about the location of the point cloud nor whether the point cloud is overlapping.
%    \item The source code demonstrating the implementation of the proposed method is provided. 
%    \\ {\color{red} upload later}


\end{frontmatter}


\section{Introduction}
\label{sec:intro}
%Unmanned Underwater vehicles (UUV) can be separated into two different categories: the remotely operated vehicle (ROV) and the autonomous underwater vehicle (AUV). In the ROV configuration, a human operator controls the underwater robot from a nearby ship through a tether. In contrast, the AUV configuration features a robot that autonomously executes a predefined mission, such as mapping the seabed in a specified area. 
3D point clouds of underwater environments are commonly acquired via active acoustic sensors, such as a Multibeam Echosounder (MBES) combined with an acoustic positioning system, or via passive optical methods, such as visual SLAM or odometry using stereo or monocular cameras, where visibility allows.

Mapping the seabed in high detail is challenging due to the surface's highly irregular and complex topology. This makes it difficult for both human operators and autonomous underwater robots to assess coverage or completeness in real-time.

Usually, a digital terrain model (DTM) or a digital surface model (DSM) is used to represent the seabed, and it can consist of a regular or irregular 2D grid with elevation data. The limitation of this geometric representation is that it cannot represent overhanging structures and intricate formations, which are quite prevalent on seabeds, especially when a detailed map is desired. 
For example, in our experimental dataset of the sunken aircraft (see \cref{fig:Heinkel}), occluded regions, such as the area beneath the aircraft's wings, were missing.  
If a DTM / DSM-based method was used, identifying missing data beneath the wing would be difficult, as a DTM represents the surface as a 2D function of elevation, $z = f(x, y)$.  
Methods that operate primarily in the 2D domain for detecting missing data therefore fail to exploit available 3D (spatial) information, particularly in scenarios involving overhangs or other complex topologies.

To address the challenges above, we propose a real-time incremental surface mesh reconstruction that maintains properties such as orientability, being a manifold, and allowing free-form structures. Simultaneously, we can identify missing data in a meaningful manner without resorting to any 3D-to-2D projection. By ``meaningful manner", we mean that missing information is identified only in regions where data should already have been acquired, rather than simply marking unexplored areas as incomplete.

Having confidence in the coverage of the gathered surface data of the seabed at the desired resolution without omitting essential information is highly advantageous. For example, by understanding the geometry of the seabed and its intrinsic optical properties, the methods described in \cite{sheinin2016next} can be potentially employed to capture clearer images with reduced light scattering. Moreover, acquiring more comprehensive geometric data generally improves the quality of post-processing, allowing the use of techniques such as photogrammetry, neural radiance fields (NeRF) \cite{mildenhall2021nerf}, Gaussian splatting \cite{kerbl20233d}, or specialized offline surface reconstruction methods for underwater point clouds \cite{campos2015surface}.

To accomplish this, we enhance a widely used surface reconstruction method that operates on point clouds with associated normal vectors, namely the Ball Pivoting Algorithm (BPA) \cite{bernardini2002ball, digne2014analysis}. The traditional BPA does not incorporate incremental point cloud processing, nor does it guarantee manifoldness or orientability. 
Many everyday physical objects occupy a well-defined volume, and their surfaces can be modeled as manifolds. In practice, manifold triangle meshes are commonly used to approximate these manifold surfaces. Later, we will show that if a surface has holes that can be closed, then the surface must be orientable.

The main contributions of this work are as follows:

%\subsection{Contribution}
\begin{itemize}
    \item An incremental BPA method that produces \textit{orientable and manifold meshes} satisfying both vertex- and edge-manifold properties. 
    \item \textit {Support for free-form surface reconstruction} beyond traditional \\DSM/DSM approaches, enabling accurate representation of overhangs, vertical walls, and other geometries with multiple $z$-values per $(x, y)$ coordinate.
    \item \textit{Versatility} in handling 3D point clouds from a variety of acquisition methods, including those with variable spatial resolution.
     \item \textit{Robustness to outliers}  in the point cloud, maintaining surface integrity in their presence.
    \item \textit{Applicability to diverse environments}, including indoor and outdoor settings.
    \item \textit{Geometric fidelity preservation}, the algorithm does not fabricate points or surfaces for visual enhancement.
    \item An \textit{online reconstruction capability} that detects mesh holes in real-time and provides actionable feedback to ROV/AUV operators about areas lacking sufficient coverage before mission completion.
    \item No assumptions on location or overlap of input point clouds.
    \item \textit{Open-source implementation}:\\ \url{https://github.com/Mauhing/Incremental-BPA}
\end{itemize}

The remainder of this article is structured as follows. \cref{sec:related} reviews the relevant literature, including real-time 3D surface reconstruction, techniques to identify missing geospatial information, and an overview of BPA. 
\cref{sec:preli} introduces the terminology and notation used throughout the article. In \cref{sec:meth:ibpa}, we describe how the original BPA can be adapted from a non-incremental to an incremental formulation.  \Cref{sec:meth:manifold} discusses the importance of manifold surface reconstruction and outlines our method. \Cref{sec:meth:orintable} focuses on achieving orientable surface reconstruction and the associated techniques. 
\Cref{sec:system} provides a comprehensive reconstruction pipeline overview, including user-configurable parameters for various operational scenarios and our method for real-time detection of missing information. 
Experimental results are presented in \cref{sec:experiment}, demonstrating the performance and robustness of our method, particularly with respect to outlier resilience. Finally, \cref{sec:concl} concludes the article and discusses potential directions for future improvement.
\section{Related Work}\label{sec:related}
\subsection{Real-time 3D Reconstruction}
\citet{newcombe2011kinectfusion} introduced KinectFusion, which uses a truncated signed distance function (TSDF) derived from \citet{curless1996volumetric} to represent the geometry for 3D reconstruction. Although they note that an isosurface can be extracted via the Marching Cubes algorithm, this step is not performed in real-time.

Building on this, \cite{oleynikova2017voxbloxa} developed Voxblox to support Micro aerial vehicle (MAV) planning with an incremental TSDF and real-time mesh extraction via the Marching Cubes algorithm, suitable for obstacle avoidance but without mentioning manifoldness and orientability. 

%Other methods such as CHISEL (\cite{Klingensmith2015chisel}) employ a dynamically spatially hashed TSDF on mobile devices, enabling real-time, large‑scale reconstruction without GPU support; however, its mesh extraction via Marching Cubes still can produce topologically inconsistent meshes that may contain non-manifold or intersecting faces, which complicate later processing steps that rely on consistent geometry.

Incremental approaches that aim to maintain manifold meshes exist and differ from volume-based methods. \citet{romanoni2015efficient, piazza2018real} propose carving-based algorithms to reconstruct manifold meshes incrementally, trading off mesh quality due to hollow artifacts (such as false surfaces like sky regions).
\citet{schops2019surfelmeshing} convert 3D surfels into meshes. However, these meshes can be non-manifold due to intersection between triangles. In contrast, our reconstruction method provides a manifold mesh without intersection.

In addition to volume- or carving-based methods, Kazhdan et al. \cite{kazhdan2006poisson} introduced Poisson surface reconstruction, which solves the Poisson equation to produce a watertight surface.
However, Poisson surface reconstruction tends to generate surfaces even in regions with sparse input data, which is undesirable for our application, since we explicitly treat the absence of data as missing information.
Yu et al. \cite{yu2019incremental} adapted Poisson surface reconstruction for incremental processing, but their method requires non-overlapping point clouds, a condition frequently violated in robotic applications. In contrast, our approach naturally handles overlapping point clouds.
Vizzo et al. \cite{vizzo2021poisson} focus primarily on SLAM applications, applying Poisson surface reconstruction to batches of oriented point clouds and subsequently merging the resulting surfaces. This merging strategy often leads to intersecting triangle meshes, an issue avoided by our method.
Ruan et al. \cite{ruan2023slamesh} also target SLAM applications, introducing Gaussian process surface interpolation for LiDAR data. However, their method does not explicitly guarantee global manifoldness or orientability and does not address missing information detection.
Niedźwiedzki et al. \cite{niedzwiedzki2023idtmm} propose a LiDAR-based incremental reconstruction approach that, during the triangle mesh expansion phase, relies on 3D-to-2D projections. This can cause triangle intersections in complex geometries, leading to non-manifold meshes.

\subsection{Missing Geo-information Detection}
%Next-best-view (NBV) planning aims to identify unobserved regions. \citet{palomeras2019autonomous,bircher2016receding} utilize occupancy-voxel models (OctoMap) and perform exhaustive random searches to identify high-information views \cite{palomeras2019autonomous,bircher2016receding}. While thorough, these random methods are computationally expensive. Yip et al. introduce deterministic hole-perimeter metrics to guide NBV without random sampling, and classify boundary holes in meshes \cite{yip2024active,yip2024robust}. However, their work is offline and does not support real-time integration with mesh reconstruction—an essential requirement for autonomous exploration.

Next-best-view (NBV) planning aims to identify unobserved regions for further exploration.   \citet{palomeras2019autonomous,bircher2016receding} utilize occupancy-voxel models (e.g., OctoMap \cite{hornung2013octomap}) and perform exhaustive random pose generation followed by virtual sensor raycasting to identify high-information gain views. In contrast, our method does not rely on random pose generation.
\citet{yip2024active,yip2024robust} propose a deterministic hole-perimeter based metrics to guide NBV planning without random sampling, and classify boundary into holes on meshes. However, the method from \cite{yip2024robust} is designed for offline use and does not support real-time integration with mesh reconstruction, needed for autonomous exploration.
In fact, most volumetric methods, including \cite{palomeras2019autonomous, bircher2016receding}, rely on exhaustive random pose generation. These methods follow a \emph{define-first-detect-later} strategy: they first define candidate poses and then detect missing information.
The widespread use of the \emph{define-first-detect-later} strategy in volumetric representations is due to inherent challenges that make it difficult for volumetric methods to adopt a detect-first-define-later approach.
In \ref{app:volVsSurf}, we analyze volumetric and surface-based methods using \(k\)-manifolds in \(\mathbb{R}^n\), and demonstrate why it is inherently difficult for volumetric approaches to adopt a \emph{detect-first-define-later} strategy.
Our method, when combined with the approach from \cite{yip2024active}, follows a \emph{detect-first-define-later} strategy: we first detect missing information and then determine the appropriate sensor poses to acquire it.

\subsection{Ball Pivoting Algorithm}
The Ball Pivoting Algorithm (BPA), introduced by \citet{bernardini2002ball}, constructs triangular meshes by pivoting a sphere across points with normals and forming triangles whose circumscribing spheres contain no other points. 
\citet{digne2014analysis} provides a detailed analysis and practical C++ implementation. BPA inherently produces edge-manifold meshes, but its traditional formulation lacks support for incremental updates, may fail to maintain orientability, and does not explicitly detect missing data. The algorithm proposed by  \cite{bernardini2002ball,digne2014analysis} operates on a point cloud with associated normal vectors. The reconstruction process begins by identifying a valid seed triangle, after which the algorithm incrementally expands the mesh by iteratively pivoting a virtual ball around boundary-edges to form new triangles. This expansion continues until no further triangles can be formed from the current boundary. At that point, the algorithm searches for a new seed triangle and resumes the expansion process. This iterative procedure continues until all vertices have been considered. 
During the expansion phase, a boundary-edge is used to evaluate candidate vertices for triangle formation. A vertex is considered suitable if three conditions are satisfied: (1) the inner product between the normal of the associated triangle candidate and the normal vector at each vertex is positive, ensuring normal consistency; (2) the triangle must accommodate a ball of a predefined radius (for example, the three points must lie on the surface of such a sphere); and (3) no other point in the cloud lies within the sphere. We refer to this as the \emph{empty ball configuration}. For further details on the algorithm and its geometric constraints, we refer to \citet{bernardini2002ball, digne2014analysis}.

%Beyond BPA, there are other manifold reconstruction methods. For example, edge-point‑based incremental triangulation builds manifold surfaces from visual SLAM point clouds \cite{niedzwiedzki2023idtmm, turn0search52}. Poisson reconstruction methods guarantee watertight manifold surfaces at scale \cite{kazhdan2006poisson}, although they do not adapt well to incremental updates or missing-data detection and are sensitive to outlier structures.\\

%In summary, while volumetric and Poisson-based methods are effective for dense modeling and planning, they do not simultaneously address three key requirements: (1) incrementality, i.e., the ability to update the mesh in real-time as new point cloud data becomes available; (2) manifoldness and orientability, which ensure that the reconstructed mesh is topologically consistent and suitable for further processing; and (3) missing-data detection, the ability to identify regions where data is lacking based on the surface geometry, rather than merely marking unobserved space. BPA-based surfel-to-mesh techniques provide high-quality interpolating surfaces with edge-manifold structure, but typically lack support for real-time updates and explicit hole inference. Our method addresses these short comings by extending BPA into an online, robust, and topology-aware reconstruction pipeline that integrates meaningful missing-information detection.

%To summarize, although many real-time reconstruction pipelines exist, none produce manifold, orientable, free-form meshes in an incremental fashion suitable for meaningful missing-information detection.

The orientable manifold triangle mesh constructed by our method is a 2D simplicial complex embedded in \(\mathbb{R}^3\) that forms a piecewise-linear 2-manifold; that is, every point (vertex) has a neighborhood homeomorphic to \(\mathbb{R}^2\).
This means that our method closely approximates a \(2\)-manifold in \(\mathbb{R}^3\), except that the reconstructed mesh is not smooth. However, if a smoothing algorithm is applied to the reconstructed mesh, it can serve as both the domain and the boundary of the domain for the Generalized Stokes Theorem:
\begin{equation}
    \int_{\submani{k}{n}} d\Phi = \int_{\psubmani{k}{n}} \Phi.
\end{equation}
In this context, the smoothed reconstructed mesh corresponds to \(\submani{2}{3}\), and its boundaries correspond to \(\psubmani{2}{3}\). For details, we refer the reader to \cite[Chapter~17.5, Theorem~4]{adams2018calculus}.

\section{Preliminaries}
\label{sec:preli}
In the subsequent sections, we establish the specialized terminology utilized in this article. (compare also \cite{yip2024robust}).
\begin{definition}[\textbf{Vertex}]
A vertex $v$ is a point in a three-dimensional Euclidean space, that is, $v \in \mathbb{R}^3$. The $i$-th vertex is denoted by $v_i$.
\end{definition}

\begin{definition}[\textbf{Edge}]
An edge $e$ is a line segment that connects two distinct vertices. If orientation is required, an edge from vertex $v_i$ to $v_j$ is denoted as $e_{ij}$.
\end{definition}

\begin{definition}[\textbf{Facet}]
A facet \(f\) is a triangle formed by interconnecting three vertices.  
If orientation is required, the facet \(f_{ijk}\) is defined by the ordered vertices \(v_i\), \(v_j\), and \(v_k\).
\end{definition}

\begin{definition}[\textbf{Boundary-edge}] \label{def:boundary-edge}
A boundary-edge $h$ is an edge adjacent to exactly one triangle. The notation $h_{ij}$ denotes a directed edge from vertex $v_i$ to $v_j$.
\end{definition}

\begin{definition}[\textbf{Full-edge}]
A full-edge is an edge that is adjacent to exactly two triangles.
\end{definition}

\begin{definition}[\textbf{Mesh}]
A triangular mesh comprises a set $\set F$ of triangles that may be connected by their common edges or vertices.
\end{definition}

\begin{figure}[htb]
   \centering
    \begin{subfigure}[t]{0.32\textwidth}
        \centering
        \includegraphics[width=\textwidth]{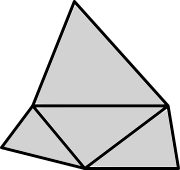}
        \caption{An edge-connected mesh.}
        \label{fig:mesh-single}
    \end{subfigure}%
    ~ 
    \begin{subfigure}[t]{0.32\textwidth}
        \centering
        \includegraphics[width=\textwidth]{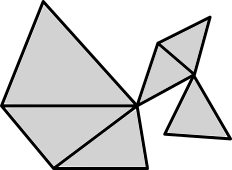}
        \caption{A vertex-connected mesh.}
        \label{fig:mesh-vertex}
    \end{subfigure}%
    \caption{Mesh illustrations. (a) Edge-connected mesh. (b) Three edge-connected meshes are connected by vertices to form a vertex-connected mesh.}
    \label{fig:mesh}
\end{figure}

\begin{definition}[\textbf{Edge-connected Mesh}]\label{def:edge-connected-mesh}
An edge-connected mesh consists of a set of triangles such that any two triangles that share a vertex $v_i$ are also connected by another vertex $v_j$, see \cref{fig:mesh-single}.
\end{definition}

\begin{definition}[\textbf{Vertex-connected Mesh}]
A vertex-connected mesh is a set of triangles where at least two triangles are connected to each other only by a single vertex and do not share any common edges, see \cref{fig:mesh-vertex}.
\end{definition}

\begin{definition}[\textbf{Edge-manifold Mesh}]\label{def:edge-mesh}
A triangle mesh is edge-manifold if and only if every edge in the mesh is adjacent to at most two triangles.
\end{definition}

\begin{definition}[\textbf{Orphan vertex}]
A vertex is an orphan vertex if and only if it is not connected to any edge or triangle.
\end{definition}

\begin{definition}[\textbf{Manifold vertex}]\label{def:manifold-vertex}
A vertex $v_i$ is a manifold vertex if and only if all following conditions are met:
\begin{enumerate}
    \item Let \(\set F_{v_i} \subseteq \set F\) represents all facets of that vertex $v_i$. \(\set F_{v_i}\) is not an empty set. This means \(\set F_{v_i} \neq \emptyset \) 
    \item If \(|\set F_{v_i}| > 1\), then for any two triangles $f_1, f_2 \in \set {F}_{v_i}$, there exists a sequence of triangles in \(\set F_{v_i}\) such that each consecutive pair shares a common edge.
\end{enumerate}
\end{definition}

\begin{definition}[\textbf{Non-manifold vertex}]\label{def:non-manifold-vertex}
A vertex $v_i$ is a non-manifold vertex if and only if all following conditions are met:
\begin{enumerate}
    \item Let \(\set F_{v_i} \subseteq \set F\) represents all facets of that vertex $v_i$. \(\set F_{v_i}\) is not an empty set. This means \(\set F_{v_i} \neq \emptyset \) 
    \item \(v_i\) is not a manifold vertex.
\end{enumerate}
\end{definition}

\begin{definition}[\textbf{Vertex-manifold Mesh}]\label{def:vertex-manifold-mesh}
A mesh is vertex-manifold if and only if all vertices in the mesh are manifold vertices.
\end{definition}

\begin{definition}[\textbf{Manifold Mesh}]
A triangle mesh is a manifold mesh if and only if it is both edge-manifold and vertex-manifold, and contains no intersecting triangles.
\end{definition}

\begin{definition}[\textbf{Boundary}]
A boundary is a closed loop composed of consecutively connected boundary-edges, denoted as \(\set b\).
\end{definition}

\begin{definition}[\textbf{Transition Edge}]\label{def:transition-edge}
Given a facet $f_{ijk}$ and an oriented edge $e_{ij}$ (from vertex $v_i$ to $v_j$), the transition edge of $e_{ij}$ is an edge that has vertex $v_j$ but not $v_i$ connected, and it is one of the edges in $f_{ijk}$. 
\end{definition}

\cref{alg:class} illustrates an efficient structure for representing geometric objects. In particular, the pointer structure enables fast topological queries to associated objects without requiring an exhaustive search, as previously implemented in \cite{digne2014analysis}. The sequence of vertices within the class facets is used to indicate the orientation of the triangular surface.

\begin{algorithm}[ht]
\DontPrintSemicolon
\SetAlgoNlRelativeSize{-1}
\SetNlSty{textbf}{}{:}
\SetKwComment{Comment}{/* }{ */}

\KwSty{Class} \textbf{Vertex} \;
\Indp
    \texttt{std::set<Edge*>} adjacentEdges \;
    \texttt{std::set<Facet*>} adjacentFacets \;
    \texttt{double} x, y, z \;
    ... \;
\Indm
\BlankLine

\KwSty{Class} \textbf{Edge} \;
\Indp
    \texttt{Vertex*} \(v_1\) \;
    \texttt{Vertex*} \(v_2\) \;
    \texttt{Facet*} \(f_1\)  \;
    \texttt{Facet*} \(f_2\)  \;
    ... \;
\Indm
\BlankLine

\KwSty{Class} \textbf{Facet} \;
\Indp
    \texttt{Vertex*} \(v_1\) \;
    \texttt{Vertex*} \(v_2\) \;
    \texttt{Vertex*} \(v_3\)  \;
    ... \;
\Indm

\caption{Class layout for Vertex, Edge, and Facet}
\label{alg:class}
\end{algorithm}

%We use curly brackets $\{...\}$ to represent a \textit{set}, following the convention from set theory; Square brackets $[...]$ denote an \textit{ordered array}, which maintains order and permits repeated elements; Angle brackets $\langle ... \rangle$ represents a \textit{cyclic array}.
\section{Methods and Algorithmic details}
This section presents the objectives of our approach and the corresponding algorithmic methods.
%First, we focus on our objective, followed by an explanation of the method to achieve it.
\label{sec:meth}
\begin{objective}\label{obj:1}
Given a stream of point clouds with associated normal vectors, the aim is to incrementally reconstruct a triangle mesh without requiring global recomputation and while preventing self-intersections between mesh triangles. 
%This objective is addressed in \cref{sec:meth:ibpa}.
%As point clouds along with their corresponding normal vectors are provided incrementally, our objective is to develop a method to gradually reconstruct a triangle mesh without re-evaluating all previously received points and ensuring that there are no intersections between the mesh triangles. This objective is accomplished in Section \cref{sec:meth:ibpa}.
\end{objective}
\begin{objective}\label{obj:2}
The triangle mesh retains both edge- and vertex-manifold properties throughout the incremental expansion process. 
%This objective is addressed in Section \cref{sec:meth:manifold}.
%The triangle mesh maintains its manifold properties, both edge-manifold and vertex-manifold, and preserves these characteristics throughout the expansion process. This objective is accomplished in Section \cref{sec:meth:manifold}.
\end{objective}
\begin{objective} \label{obj:3}
%The mesh must maintain orientability, ensuring that all triangle normals are consistently directed. This objective is addressed in Section~\cref{sec:meth:orintable}.
The mesh is orientable throughout the incremental expansion process. 
%This objective is addressed in \cref{sec:meth:orintable}.
\end{objective}
These objectives are addressed subsequently in the following \cref{sec:meth:ibpa}, \cref{sec:meth:manifold} and \cref{sec:meth:orintable}.

\begin{figure}[htbp]
    \centering
    \begin{subfigure}[b]{0.48\textwidth}
        \centering
        \includegraphics[width=\textwidth]{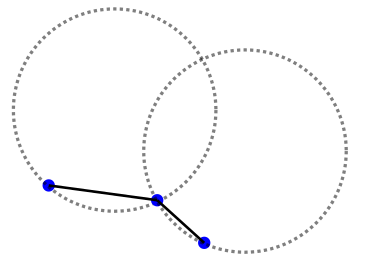}
        \caption{Surface reconstruction at time $t$.}
        \label{fig:emptyConfig-1}
    \end{subfigure}
    \hfill
    \begin{subfigure}[b]{0.48\textwidth}
        \centering
        \includegraphics[width=\textwidth]{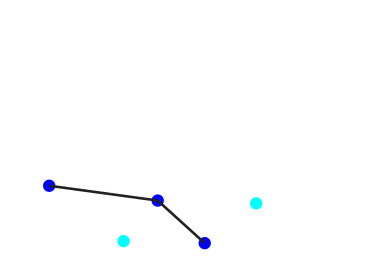}
        \caption{Incoming point cloud at time $t+1$, shown in light blue.}
        \label{fig:emptyConfig-2}
    \end{subfigure}
    \vfill
    \begin{subfigure}[b]{0.48\textwidth}
        \centering
        \includegraphics[width=\textwidth]{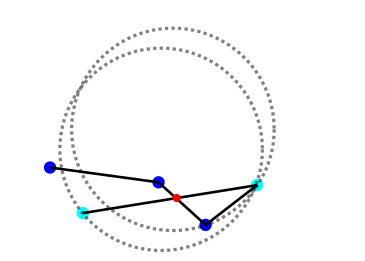}
        \caption{Reconstruct further on the new point cloud and lead to surface intersection, shown in the red point.}
        \label{fig:emptyConfig-3}
    \end{subfigure}
    \hfill
    \begin{subfigure}[b]{0.48\textwidth}
        \centering
        \includegraphics[width=\textwidth]{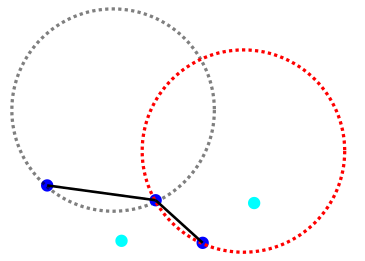}
        \caption{Shown the source of the problem. Violation of empty ball configuration.}
        \label{fig:emptyConfig-4}
    \end{subfigure}
\caption{The first three figures (a, b, and c) illustrate a failure mode of applying the original BPA algorithm incrementally. As new point clouds are added (b), the algorithm may produce surface intersections (c). The root cause is a violation of the empty ball configuration, shown in (d).
}
    \label{fig:emptyConfig}
\end{figure}

\subsection{Transforming BPA into an Incremental Algorithm}\label{sec:meth:ibpa}
To adapt the original Ball Pivoting Algorithm for incremental reconstruction, two primary challenges must be addressed.
First, the spatial data structure, originally a static octree used to accelerate neighbor searches, must be made dynamic to accommodate the continual insertion of new points during reconstruction.
Second, each newly added point must be checked against the existing mesh to determine whether it violates the empty ball configuration. If a new point lies within the circumsphere of any existing triangle, see \cref{fig:emptyConfig}, that triangle must be removed to preserve surface validity, see \cref{fig:emptyConfig-solution}. However, the vertices of the removed triangle are retained for future use.
%Second, for every new incoming point, we need to check whether it is inside any ball of any triangle surface. If it is, we remove the triangle but do not remove any vertex.

\subsubsection{Expandable Octree Data Structure}
\label{sec:meth:octree}
\begin{figure}[htbp]
    \begin{subfigure}[b]{0.48\textwidth}
        \centering
        \includegraphics[width=\textwidth]{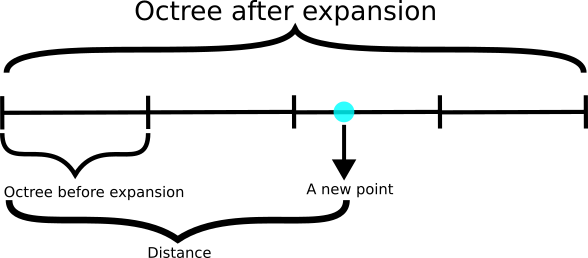}
        \caption{Expanding direction to the right.}
        \label{fig:octree-expansion-r}
    \end{subfigure}
    \hfill
    \centering
    \begin{subfigure}[b]{0.48\textwidth}
        \centering
        \includegraphics[width=\textwidth]{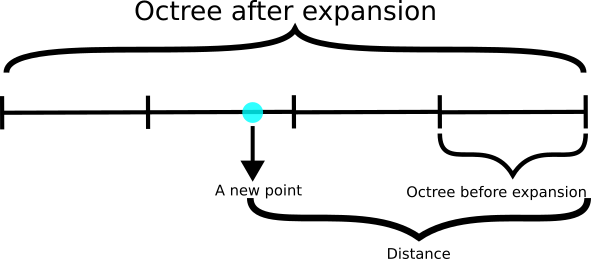}
        \caption{Expanding direction to the left.}
        \label{fig:octree-expansion-l}
    \end{subfigure}
    \caption{1D illustration for two cases of octree expansion. When a new point lies outside the current octree bounds, the structure expands in the appropriate direction.}
    \label{fig:octree-expansion}
\end{figure}

When a new vertex lies outside the current bounds of the octree, we must expand the structure to encompass it. For illustration, consider expansion along a single axis, for instance, the x-axis.
There are two scenarios to consider. In one scenario, a vertex is located on the right side of the current octree, as depicted in\cref{fig:octree-expansion-r}. In the other scenario, a vertex is located on the left side of the current octree, as shown in  \cref{fig:octree-expansion-l}. 
Both scenarios are quite similar, although in the second scenario it is necessary to redefine the origin of the octree.
Let us study the 1D case of \cref{fig:octree-expansion-r}. We aim to determine the required increase in the octree level. Let the distance be denoted by $d$, and let $l_b$ be the length of the octree before expansion. The required increment $\Delta n$ can then be computed using the following equation:
\begin{equation}
    d \leq \min_{\Delta n} (l_b 2^{\Delta n}), \quad \Delta n \in \mathbb{N}.
\end{equation}
This implies that
\begin{equation}
    \Delta n = \left\lceil \log_2 \frac{d}{l_b} \right\rceil.
    \label{eq:octree-expansion-delta-n}
\end{equation}
The total level of the octree after expansion is $n_a = n_b + \Delta n$, where $n_b$ denotes the level before expansion and $n_a$ denotes the level after expansion. The total length of the octree after expansion is $l_a = R \cdot 2^{n_a}$, where $R$ is the radius of the ball and $l_a$ is the length of the octree after expansion.

For all the vertices that are inside the octree before expansion, their 1D locational code after expansion is given by:
\begin{equation}
    \underbrace{000\ldots000}_{\Delta n}\underbrace{x_{n_b} \ldots x_2 x_1}_{n_b}, \quad x_i \in \{0, 1\}.
\end{equation}
This corresponds to simply appending $\Delta n$ zeros to the left of the locational code before expansion.  
The case in \cref{fig:octree-expansion-l} is identical to that in \cref{fig:octree-expansion-r}, except for two differences.  
First, the locational code of all the vertices that are inside the octree before expansion is given by:
\begin{equation}
    \underbrace{111\ldots111}_{\Delta n}\underbrace{x_{n_b} \ldots x_2 x_1}_{n_b}, \quad x_i \in \{0, 1\}.
\end{equation}
Second, the starting point of the octree is redefined to align with the leftmost of the x-axis. For the other two axes, the behavior remains unchanged. However, to maintain a consistent octree level $n_a$ across all axes, we also expand the other axes even if it is not strictly necessary. The final level is chosen as $n_a = \max(n_{a,x}, n_{a,y}, n_{a,z})$. 
For any axis that does not require expansion but is still expanded for consistency, we apply right-sided expansion, as illustrated in \cref{fig:octree-expansion-r}. This ensures a consistent octree level $( n_a )$ across all axes.
%For any axis that does not require expansion but is still expanded for consistency, we apply right-sided expansion, as illustrated in Figure~\cref{fig:octree-expansion-r}.

The following example illustrates the octree expansion process.
We assume that the starting point is at $(0, 0, 0)$ and the initial octree length is $l_b = 5$. This means that if all vertices lie within the bounding box from $(0, 0, 0)$ to $(5, 5, 5)$, no expansion is required. However, we consider a vertex located at $(13, -3, 2)$.
\begin{itemize}
    \item Along the x-axis, we must expand twice, so that $\Delta n_x = 2$.
    \item Along the y-axis, we need to expand once, so that $\Delta n_y = 1$.
    \item Along the z-axis, no expansion is needed, so that $\Delta n_z = 0$.
\end{itemize}
Since we want to use the same octree level for all axes, we take the maximum expansion: $\Delta n = \max(\Delta n_x, \Delta n_y, \Delta n_z) = 2$. Therefore, all axes are expanded by two levels.
\begin{itemize}
    \item For the x-axis, we expand the octree to the right.
    \item For the y-axis, we expand to the left.
    \item For the z-axis, although no expansion is needed, we still expand to the right for consistency.
\end{itemize}
The new starting point becomes $(0, -15, 0)$ and the new octree length is $l_a = 20$, since \( 5 \times 2^2 = 20 \). The expansion along y-axis required left-sided growth with two levels, the starting point of y-axis shifts to \(-15\) to maintain the bounding box. As a result, the new octree now covers the space from $(0, -15, 0)$ to $(20, 5, 20)$.

\subsubsection{Ensure Empty Ball Configuration}
\label{sec:meth:emptyball}
In the original BPA algorithm, ensuring the empty ball configuration is straightforward, as all vertices are known in advance and no new vertices will be added. However, in incremental BPA, we must ensure the empty ball configuration for every newly arriving vertex.
To achieve this, for each incoming vertex, we check all existing triangle-associated balls. If the vertex lies inside any of these balls, we remove the corresponding triangle but do not remove any vertices.
Checking all balls for every new vertex can be computationally expensive. To address this, we introduce an additional octree data structure to store the centers of all existing balls, in addition to the octree used for incoming vertices. However, the spatial length of the octree can only expand or remain the same during updates, it cannot shrink.

This enables efficient neighbor search: for each new vertex, we only need to examine balls whose centers distance smaller than radius of the ball.

\begin{figure}[htbp]
    \centering
    \begin{subfigure}[b]{0.32\textwidth}
        \centering
        \includegraphics[width=\textwidth]{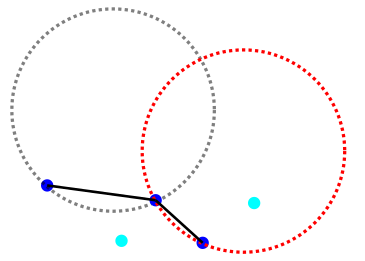}
        \caption{}
        \label{fig:emptyConfig-solution-a}
    \end{subfigure}
    \hfill
    \begin{subfigure}[b]{0.32\textwidth}
        \centering
        \includegraphics[width=\textwidth]{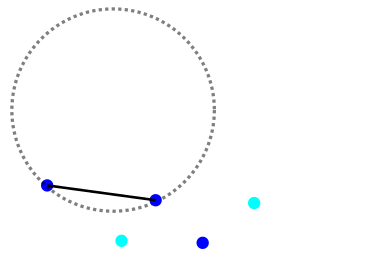}
        \caption{}
        \label{fig:emptyConfig-solution-b}
    \end{subfigure}
    \hfill
    \begin{subfigure}[b]{0.32\textwidth}
        \centering
        \includegraphics[width=\textwidth]{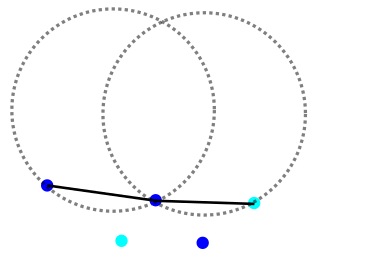}
        \caption{}
        \label{fig:emptyConfig-solution-c}
    \end{subfigure}
    \caption{The incremental BPA process. (a) shows the initial surface reconstruction; (b) illustrates how the affected triangles are identified and removed upon the arrival of a new vertex; and (c) presents the final surface after incorporating the new vertex.}
    \label{fig:emptyConfig-solution}
\end{figure}

Our solution to maintaining the empty ball configuration is to delete any surface triangle whose associated ball contains a newly incoming vertex. Importantly, no vertices are deleted in this process. \cref{fig:emptyConfig-solution} illustrates this procedure.
In \cref{fig:emptyConfig-solution-a}, we show the initial surface reconstruction with an incoming vertex that is inside a ball.
In \cref{fig:emptyConfig-solution-b}, the surface is removed and all vertices are kept.
In \cref{fig:emptyConfig-solution-c}, surface reconstruction is performed such that there is no vertex inside any new ball associated with any surface.

\subsubsection{Incremental Surface Expansion}
\label{sec:meth:incremental}
Having devised a solution using an expandable octree data structure and ensured the empty ball configuration, we can now incrementally build the surface without introducing any facet intersections.  
At this stage, we have successfully achieved our \cref{obj:1}.  
However, the issue of non-manifold vertices still remains and must be addressed.

\subsection{Problem with Non-manifold Vertex}\label{sec:meth:manifold}
The original BPA ensures that the triangle mesh is edge-manifold, but it does not guarantee that the mesh is vertex-manifold. Our goal is to ensure that our method produces a mesh that is both edge-manifold and vertex-manifold. We begin by distinguishing between two types of non-manifold vertices. Then, we explain why it is important for our method to enforce vertex-manifoldness.

\begin{figure}[htbp]
    \centering
    \begin{subfigure}[b]{0.48\textwidth}
        \centering
        \includegraphics[width=\textwidth]{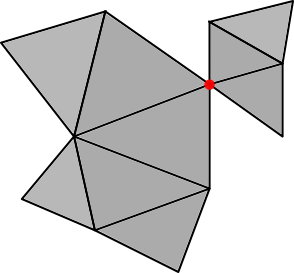}
        \caption{$0$-disk non-manifold vertex.}
        \label{fig:nonmanifold-tyeps-0-disk}
    \end{subfigure}
    \hfill
    \begin{subfigure}[b]{0.48\textwidth}
        \centering
        \includegraphics[width=\textwidth]{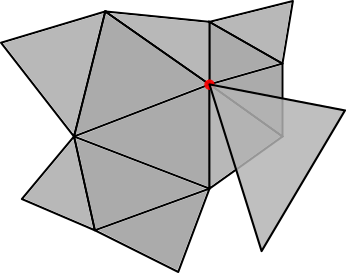}
        \caption{$1$-disk non-manifold vertex.}
        \label{fig:nonmanifold-types-1-disk}
    \end{subfigure}
    \caption{Two types of non-manifold vertices. Left: the $0$-disk non-manifold vertex. Right: the $1$-disk non-manifold vertex. Both types can occur in the BPA algorithm of \cite{digne2014analysis}.}
    \label{fig:nonmanifold-types}
\end{figure}

\cref{fig:nonmanifold-types} illustrates two different types of non-manifold vertices.  
We define the configuration on the left as a \textbf{$0$-disk non-manifold vertex}, as there is no closed loop of surfaces incident to the vertex.  
In contrast, the configuration on the right contains a closed loop of surfaces around the vertex, but includes a boundary-edge. We define this as a \textbf{$1$-disk non-manifold vertex}.
In the following, we provide detailed definitions of the two types of non-manifold vertices.
\begin{definition}[\textbf{\(1\)-disk Non-manifold Vertex}]
A vertex \(v_i\) is a \(1\)-disk non-manifold vertex if and only if it satisfies all the following conditions:
\begin{enumerate}
    \item \(v_i\) is a non-manifold vertex.
    \item Given the set of adjacent facets \(\set{F}_{v_i}\), there exists \textbf{one and only one} subset \(\set{F}_\epsilon \subseteq \set{F}_{v_i}\) such that:
    For any facet \(f \in \set{F}_\epsilon\), we can arbitrarily choose a starting facet and one of its edges incident to \(v_i\). Using this edge and the vertex \(v_i\), we identify the corresponding transition edge. This transition edge is shared by exactly one other facet in \(\set{F}_\epsilon\), which we then visit. By repeatedly following transition edges in this manner, we eventually return to the starting facet \(f\), having visited every facet in \(\set{F}_\epsilon\) exactly once.
\end{enumerate}
\end{definition}
\begin{definition}[\textbf{\(0\)-disk Non-manifold Vertex}]
A vertex \(v_i\) is a \(0\)-disk non-manifold vertex if and only if it satisfies all the following conditions:
\begin{enumerate}
    \item \(v_i\) is a non-manifold vertex.
    \item Given the set of adjacent facets \(\set{F}_{v_i}\), there does \textbf{not} exist any subset \(\set{F}_\epsilon \subseteq \set{F}_{v_i}\) such that:
    For any facet \(f \in \set{F}_\epsilon\), we can arbitrarily choose a starting facet and one of its edges incident to \(v_i\). Using this edge and the vertex \(v_i\), we identify the corresponding transition edge. This transition edge is shared by exactly one other facet in \(\set{F}_\epsilon\), which we then visit. By repeatedly following transition edges in this manner, we eventually return to the starting facet \(f\), having visited every facet in \(\set{F}_\epsilon\) exactly once.
\end{enumerate}
\end{definition}

We aim to avoid both types of non-manifold vertices so that our method incrementally reconstructs a manifold surface.  
There is one additional type of non-manifold vertex, known as an \textbf{$n$-disk non-manifold vertex} for $n > 1$, where the local neighborhood consists of multiple disconnected surface patches, each forming its own disk.
However, this case does not occur in the BPA framework; we omit the proof here. 
We will first discuss why we want to avoid 1-disk non-manifold vertices, how to detect them, and how to remove them. Then, we will do the same for 0-disk non-manifold vertices.

\subsubsection{1-disk Non-manifold Vertex}
\begin{figure}[htbp]
    \centering
    \begin{subfigure}[b]{0.48\textwidth}
    \centering
    \includegraphics[width=\textwidth]{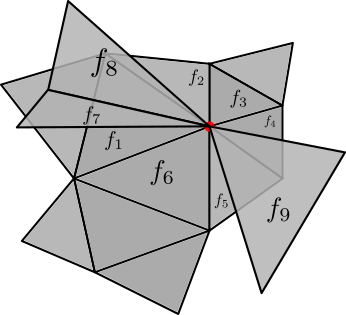}
    \caption{Before}
    \label{fig:non-manifold-vertex-one-disk-problem}
    \end{subfigure}
    \hfill
    \begin{subfigure}[b]{0.48\textwidth}
        \centering
        \includegraphics[width=\textwidth]{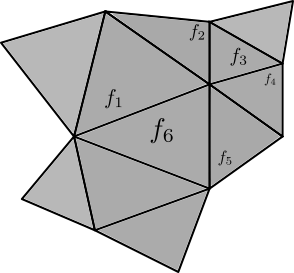}
        \caption{After}
        \label{fig:non-manifold-vertex-one-disk-solution}
    \end{subfigure}
    \caption{An auxiliary figure to explain the function \texttt{getListOfConnectedFacetsSet}. In (a), the red vertex is a $1$-disk non-manifold vertex. In this case, we obtain $\set M = [\{f_1, \ldots, f_6\}, \{f_7, f_8\}, \{f_9\}]$. \cref{alg:remove-all-redundant-one-disk-facets} will remove $\{f_7, f_8, f_9\}$ from (a), and the result is shown in (b), where the red vertex is no longer a $1$-disk non-manifold vertex.}
    \label{fig:non-manifold-vertex-one-disk}
\end{figure}

We aim to avoid $1$-disk non-manifold vertices, as they do not correspond to any physically plausible surface, as illustrated in \cref{fig:nonmanifold-types-1-disk}. Such configurations create an ambiguous local surface topology, violating manifoldness by introducing an interior boundary where none should exist. 
If left unaddressed, this type of non-manifold configuration may grow progressively larger during surface reconstruction.
\cref{alg:remove-all-redundant-one-disk-facets} demonstrates how to detect and remove redundant disconnected facets around $1$-disk non-manifold vertices immediately after each batch reconstruction, using \cref{fig:non-manifold-vertex-one-disk} as an illustrative example.

\begin{algorithm}[ht]
    \caption{$\text{removeAllRedundantOneDiskFacets}(\set F_{\text{fresh}}, \set E_{\text{boundary}})$.}
    \label{alg:remove-all-redundant-one-disk-facets}
    $\set V \gets \text{getAllVertices}(\set F_{\text{fresh}}) \cap \text{getAllVertices}(\set E_{\text{boundary}})$\; 
    $\set F_{\text{delete}} \gets \emptyset$\;
    \For{$ v \in \set V $ }
    {
        $\set M \gets \text{getListOfConnectedFacetsSet}(v)$\;
        \If{$\set M$ has 1-disk facets}
        {
            $\set F_{\text{1-disk}} \gets \text{getTheOneDiskFacets}(\set M)$\;
            $\set M_{\text{redundant}} \gets \set M \setminus \set F_{\text{1-disk}}$\;
            $\set F_{\text{redundant}} \gets \text{turnListOfSetInToSet}(\set M_{\text{redundant}})$\;
            $\set F_{\text{delete}} \gets \set F_{\text{delete}} \cup \set F_{\text{redundant}}$
        }
    }
    \For{$f \in \set F_{\text{delete}}$}
    {
        $\text{deleteFacet}(f,\, \text{keep\_vertex} = False)$
    }
\end{algorithm}

The following is the explanation of the \cref{alg:remove-all-redundant-one-disk-facets}.
\begin{itemize}
    \item $\set F_{\text{fresh}}$: the set of newly added facets in the current batch. $\set E_{\text{boundary}}$: the set of boundary-edges of the entire mesh, i.e., edges connected to only one triangle facet (not two).
    \item Line 1: Extract the vertices from $\set F_{\text{fresh}}$ and $\set E_{\text{boundary}}$, compute their intersection, and store the result in the set $\set V$.
    Since \(\set V\) is a set its vertices are, by definition, unique.
    \item Line 4: The function \texttt{getListOfConnectedFacetsSet} identifies all connected sets of facets around a vertex. In the case of  \cref{fig:non-manifold-vertex-one-disk}, it will be 
    $\set M = [ \{f_1, \ldots, f_6\}, \{f_7, f_8\}, \{f_9\} ]$
    \item Line 5: Determine whether any subset in $\set M$ forms a $1$-disk configuration. In \cref{fig:non-manifold-vertex-one-disk}, the set $\{f_1, \ldots, f_6\}$ forms a $1$-disk.  
    Since the original BPA ensures edge-manifold, it can be checked whether the facets form a single connected disk around the vertex with exactly one boundary-edge.
    \item Line 6: Extract the $1$-disk facets and store them in $\set F_{\text{1-disk}}$.  
    In the example of \cref{fig:non-manifold-vertex-one-disk}: $\set F_{\text{1-disk}} = \{f_1, \ldots, f_6\}$
    \item Line 7: Extract all remaining facet sets as redundant and store them in $\set M_{\text{redundant}}$.  
    In the example of \cref{fig:non-manifold-vertex-one-disk}: $\set M_{\text{redundant}} = [\{f_7, f_8\}, \{f_9\}]$
    \item Line 8: Flatten $\set M_{\text{redundant}}$ into a single set $\set F_{\text{redundant}}$.  
    In the example of \cref{fig:non-manifold-vertex-one-disk}: $\set F_{\text{redundant}} = \{f_7, f_8, f_9\}$
    \item Line 9: Delete all facets in $\set F_{\text{redundant}}$.  
    In the example of \cref{fig:non-manifold-vertex-one-disk}, we remove $\{f_7, f_8, f_9\}$.  
    If a vertex is no longer connected to any facet afterward, it is also removed from the octree.
\end{itemize}
Note that we only need to consider $\set F_{\text{fresh}}$ rather than the entire triangle mesh. This is crucial for real-time performance, as the size of $\set F_{\text{fresh}}$ does not grow with the overall mesh.
Importantly, as shown in \cref{fig:non-manifold-vertex-one-disk-solution}, the valid $1$-disk facets $\set F_{\text{1-disk}}$ are preserved while redundant patches are removed.

\subsubsection{0-disk Non-manifold Vertex}
\begin{figure}[htbp]
    \centering
    \begin{subfigure}[b]{0.48\textwidth}
        \centering
        \includegraphics[width=\textwidth]{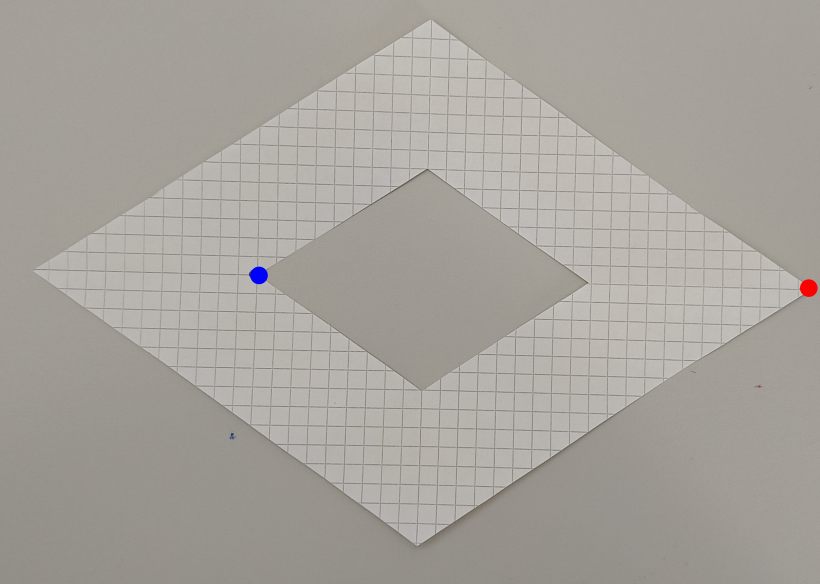}
        \caption{}
        \label{fig:0-disk-nonmanifold-a}
    \end{subfigure}
    \hfill
    \begin{subfigure}[b]{0.48\textwidth}
        \centering
        \includegraphics[width=\textwidth]{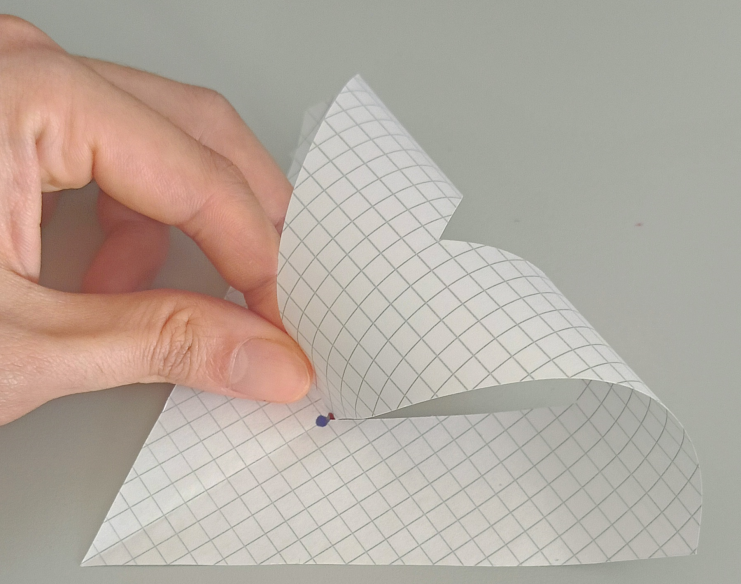}
        \caption{}
        \label{fig:0-disk-nonmanifold-b}
    \end{subfigure}
    \caption{In (a), the hole can be easily detected and is fillable. In (b), if the red vertex touches the blue vertex, it becomes a single non-manifold vertex. In this case, the hole cannot be filled; attempting to do so will result in a $1$-disk non-manifold vertex.}
    \label{fig:0-disk-nonmanifold}
\end{figure}

There are two reasons why we want to avoid $0$-disk non-manifold vertices.  
First, they do not represent physically plausible surfaces, for the same reason we avoid $1$-disk non-manifold vertices.  
Second, in certain mesh configurations, they may result in holes that cannot be filled, as illustrated in \cref{fig:0-disk-nonmanifold}.  
In \cref{fig:0-disk-nonmanifold-b}, imagine lifting the red vertex and folding it toward the blue vertex so that both coincide. In this case, the hole cannot be filled: trying to do so will create a $1$-disk non-manifold vertex, another undesired configuration. 
%Such configurations can arise if new data points are added near existing boundary vertices, effectively collapsing the local geometry. 
Moreover, such holes are challenging to fill in practice, especially for underwater robots performing real-time reconstruction tasks.

\begin{figure}[htbp]
    \centering
    \begin{subfigure}[b]{0.32\textwidth}
        \centering
        \includegraphics[width=\textwidth]{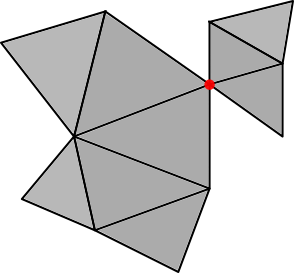}
        \caption{}
        \label{fig:non-manifold-vertex-0-disk-1}
    \end{subfigure}
    \hfill
    \begin{subfigure}[b]{0.32\textwidth}
        \centering
        \includegraphics[width=\textwidth]{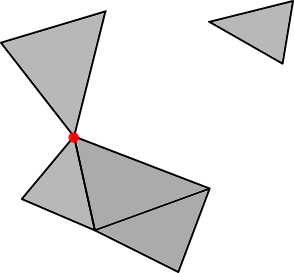}
        \caption{}
        \label{fig:non-manifold-vertex-0-disk-2}
    \end{subfigure}
    \hfill
    \begin{subfigure}[b]{0.22\textwidth}
        \centering
        \includegraphics[width=\textwidth]{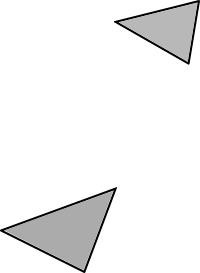}
        \caption{}
        \label{fig:non-manifold-vertex-0-disk-3}
    \end{subfigure}
    \caption{Deleting facets that contain a $0$-disk non-manifold vertex may also introduce a new $0$-disk non-manifold vertex. This is why a \texttt{while} loop is necessary in \cref{alg:remove-zero-disk-facets}.}
    \label{fig:non-manifold-vertex-0-disk-remove}
\end{figure}

To detect and remove facets that contain a $0$-disk non-manifold vertex, we use \cref{alg:remove-zero-disk-facets}. The following provides an explanation of \cref{alg:remove-zero-disk-facets}.
\begin{algorithm}[ht]
    \caption{$\text{removeZeroDiskFacets}(\set E_{\text{boundary}}, \text{keepVertex} = \text{false})$.}
    \label{alg:remove-zero-disk-facets}
    \While{true}
    {
        $\set V \gets \text{getRepeatedVertices}(\set E_{\text{boundary}})$\;
        $\set F_{\text{delete}} \gets \emptyset$\;
        \For{$v \in \set V$}
        {
            $\set F_{\text{delete}} \gets \set F_{\text{delete}} \cup \text{getFacets}(v)$
        }
        \For{$f \in \set F_{\text{delete}}$}
        {
            $\text{deleteFacet}(f,\, \text{keepVertex})$
        }
        \If{$\set F_{\text{delete}}$ is empty}
        {
            \Break\;
        }
    }
\end{algorithm}
\begin{itemize}
    \item Line 2: Identify all boundary vertices with degree larger than two (i.e., repeated in the boundary-edge list), which indicates a $0$-disk non-manifold vertex. Since one edge’s end is another’s start, the minimum repetition count is $2$. If a vertex appears more than twice, it indicates that more than two boundary-edges are connected to the same vertex.
    \item Line 3: Initialize the set of facets to delete. Note, this is a set structure, implying it automatically avoids duplicate facets.
    \item Line 5: For each repeated vertex, retrieve all facets associated with it.
    \item Line 8: Delete the retrieved facets. For any vertex that becomes orphaned (i.e., no longer connected to any facet) during this process, we also remove the vertex.
\end{itemize}
The \texttt{while} loop is necessary because deleting these facets can modify the local boundary structure, potentially creating new $0$-disk non-manifold vertices in subsequent iterations.  
Therefore, it is necessary to repeatedly check and remove such configurations. See \cref{fig:non-manifold-vertex-0-disk-remove} for an illustrative example.  
At this stage, we have successfully achieved \cref{obj:2}.

\subsection{Ensure Orientable Triangle Mesh}\label{sec:meth:orintable}
\begin{figure}[htbp]
    \centering
    \begin{subfigure}[b]{0.32\textwidth}
        \centering
        \includegraphics[width=\textwidth]{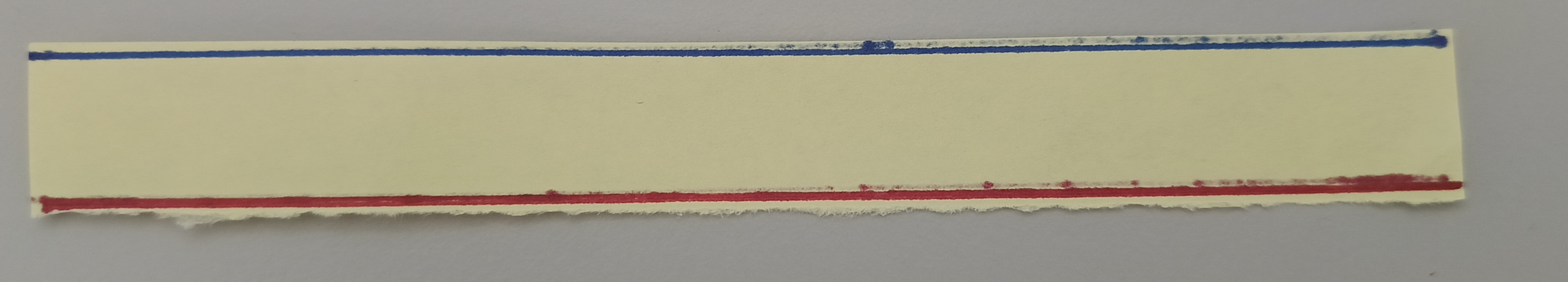}
        \caption{}
        \label{fig:orientation-1}
    \end{subfigure}
    \hfill
    \begin{subfigure}[b]{0.32\textwidth}
        \centering
        \includegraphics[width=\textwidth]{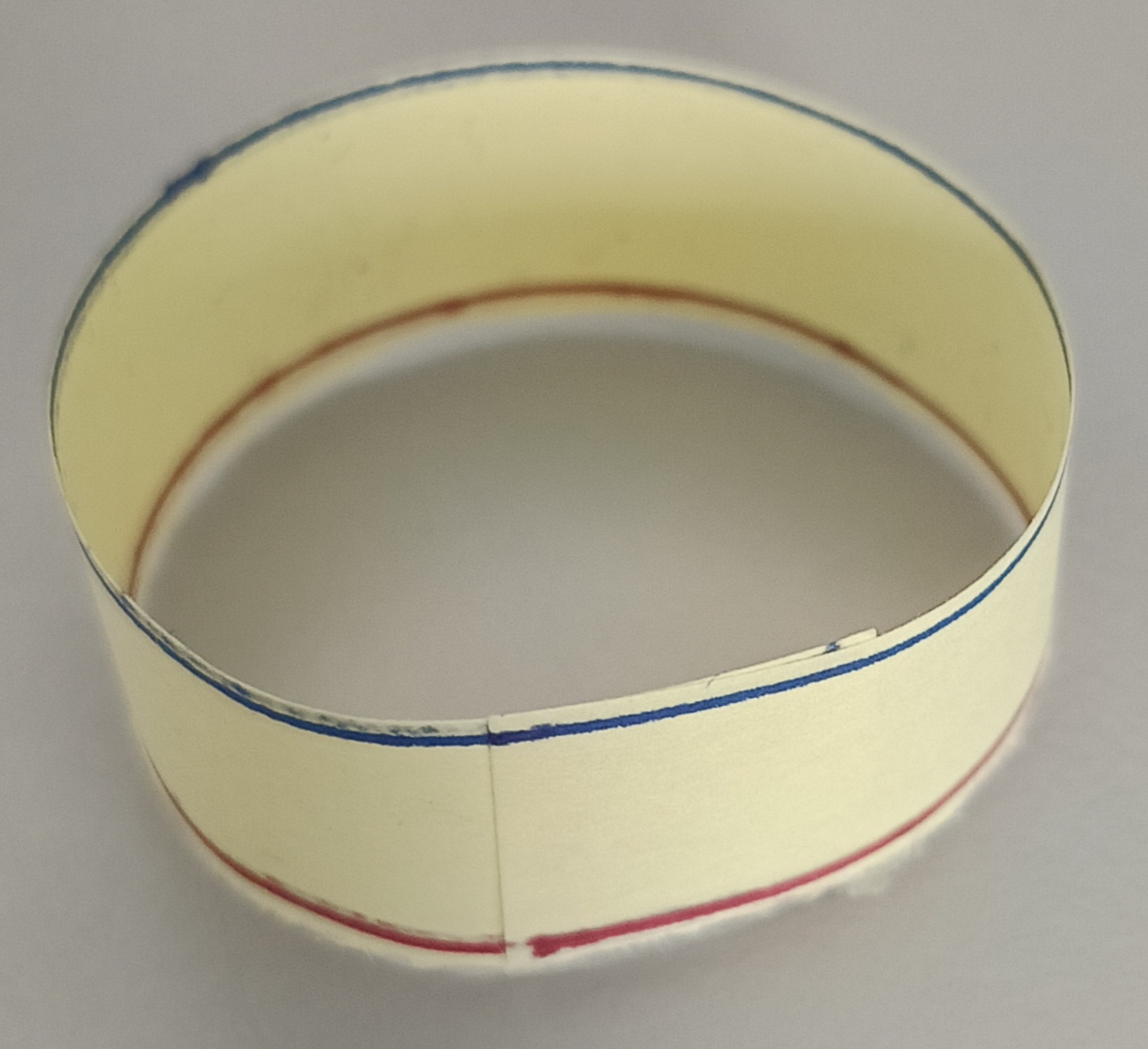}
        \caption{}
        \label{fig:orientation-2}
    \end{subfigure}
    \hfill
    \begin{subfigure}[b]{0.32\textwidth}
        \centering
        \includegraphics[width=\textwidth]{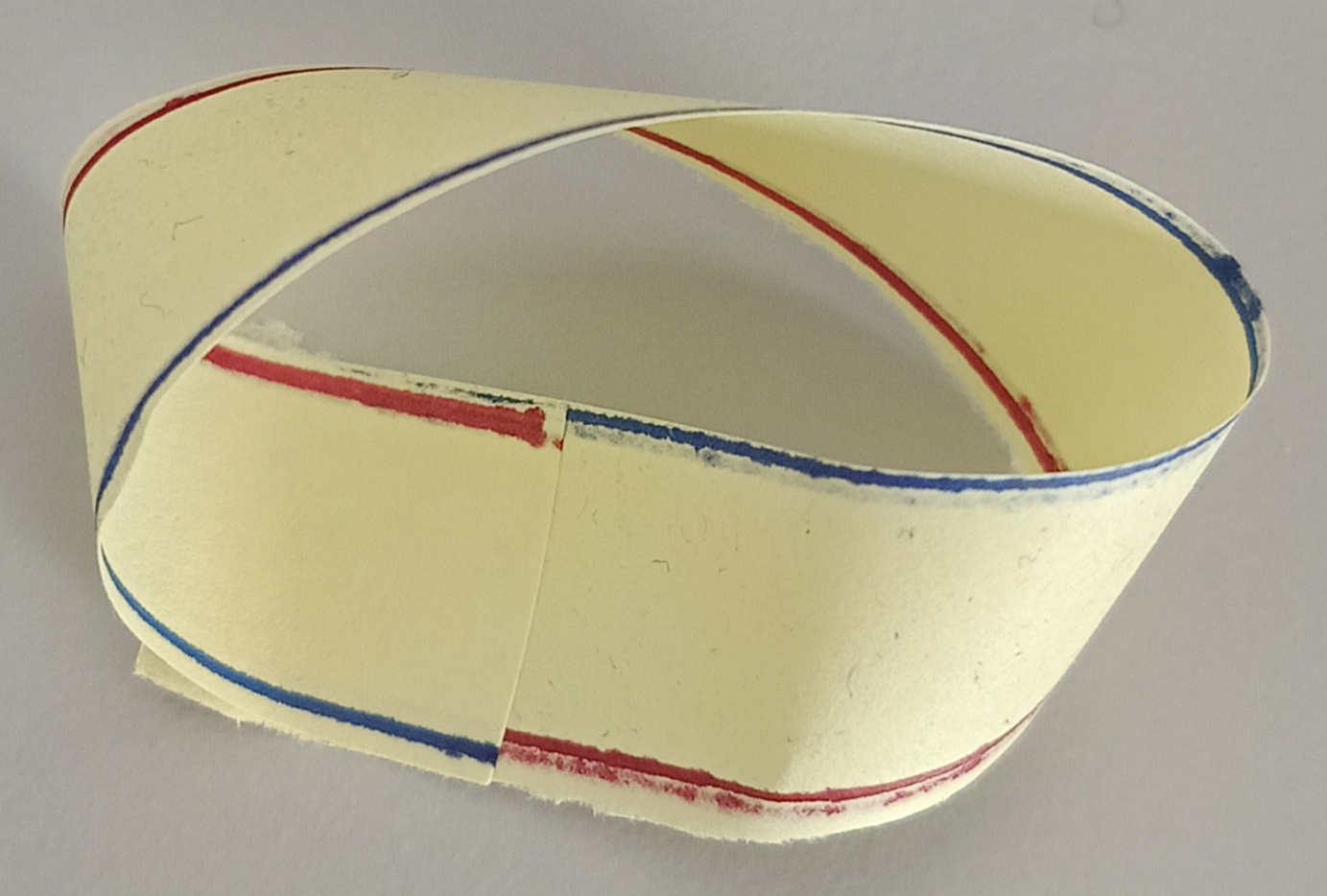}
        \caption{}
        \label{fig:orientation-3}
    \end{subfigure}
    \caption{In (a), a strip is shown with two edges marked in blue and red. In (b), the two ends of the strip are connected, forming two boundaries, blue and red, which also define two holes. Filling these holes may results in a cylinder. In (c), the strip is first twisted before connecting the blue and red edges in an interchanged manner. This creates a single boundary and results in the well-known M\"obius strip.}
    \label{fig:orentation-why}
\end{figure}

We want the reconstructed mesh to be orientable. If a configuration such as a M\"obius strip arises, see \cref{fig:orentation-why}, it would be non-orientable and therefore invalid for our purposes. %which is non-orientable, see \cref{fig:orentation-why}.

\begin{quote}
\textit{\enquote{We call a surface orientable if it does not contain a M\"obius strip; we call it non-orientable if it does contain a M\"obius strip.}}~\citet{zeeman1966introduction}
\end{quote}
Although \citet{zeeman1966introduction} uses the phrase ``we call'', it is a widely accepted consensus in topology that a surface is non-orientable if and only if it contains a M\"obius strip as a subset.

Since a M\"obius strip (or M\"obius band) has only one boundary, it is possible to attempt to ``fill'' it using another surface with a single boundary. If one glues two M\"obius strips together along their boundaries, the result is a Klein bottle (\cite{sequin2012moebius}). The Klein bottle is a \(2\)-manifold in \(\mathbb{R}^4\) without self-intersections. However, when it is represented in \(\mathbb{R}^3\), it necessarily self-intersects, and therefore cannot be represented as a valid manifold mesh in three dimensions. Moreover, it is non-orientable.

For example, if we glue a M\"obius strip to a disk (an orientable surface with one boundary), the result is a topological projective plane, which is also non-orientable.
By a classical result due to \citet{zeeman1966introduction}, any closed, connected, non-orientable surface in 3D must self-intersect. Consequently, any attempt to extend a surface containing a M\"obius strip to a closed surface in three dimensions necessarily introduces self-intersections. 

% Fromthe corollary of \citet{zeeman1966introduction} :
%\begin{quote}
%\textit{\enquote{Corollary: Any non-%orientable\footnote{We believe Zeeman forgot the term %`non-orientable'. Otherwise, a simple counter example %such as a sphere would invalidate the %corollary.}closed connected surface in 3-dimensions %must have self-%intersections}}~\citet{zeeman1966introduction}
%\end{quote}
% From the corollary, we see that if we attempt to fill a surface that contains a M\"obius strip as a subset, the result will always be a closed surface with self-intersections. This is precisely why we aim to avoid non-orientable reconstructions.

\begin{figure}
    \centering
    \includegraphics[width=0.5\linewidth]{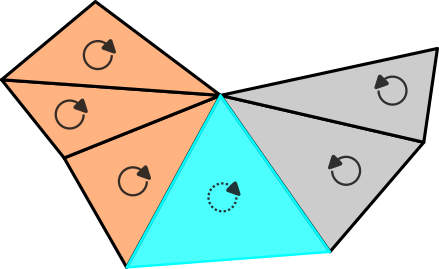}
    \caption{The orange facets are oriented oppositely to the gray facets. The suggested facet, shown in light blue, is rejected because its orientation differs from that of the neighboring gray facet.}
    \label{fig:reject-orientation}
\end{figure}

This is precisely why we aim to avoid non-orientable reconstructions and to prevent this issue, we enforce orientation consistency whenever a new triangle is added. Specifically, we ensure that the new triangle shares the same orientation with all its neighboring triangles. If an edge is on the boundary (i.e., has no adjacent triangle yet), no orientation check is needed for that edge. A triangle is only accepted if it is consistent in orientation with all its neighbors (see \cref{fig:reject-orientation}).

Since each triangle has exactly three edges, we need to check at most three neighboring triangles. This is computationally efficient, as each edge stores a reference to its adjacent facets. At this stage, we have successfully achieved \cref{obj:3}.

%\begin{figure}
%    \centering
%    \includegraphics[width=0.8\linewidth]{Figures/klein-bottle.png}
%    \caption{The klein bottle in 3D. Source: \cite{mathstackKleinBottle}
%    \label{fig:klein-bottle}
%\end{figure}

\section{Reconstruction Pipeline Overview and Additional Features} \label{sec:system}
\begin{figure}[htbp]
    \centering
    \includegraphics[width=0.5\textwidth]{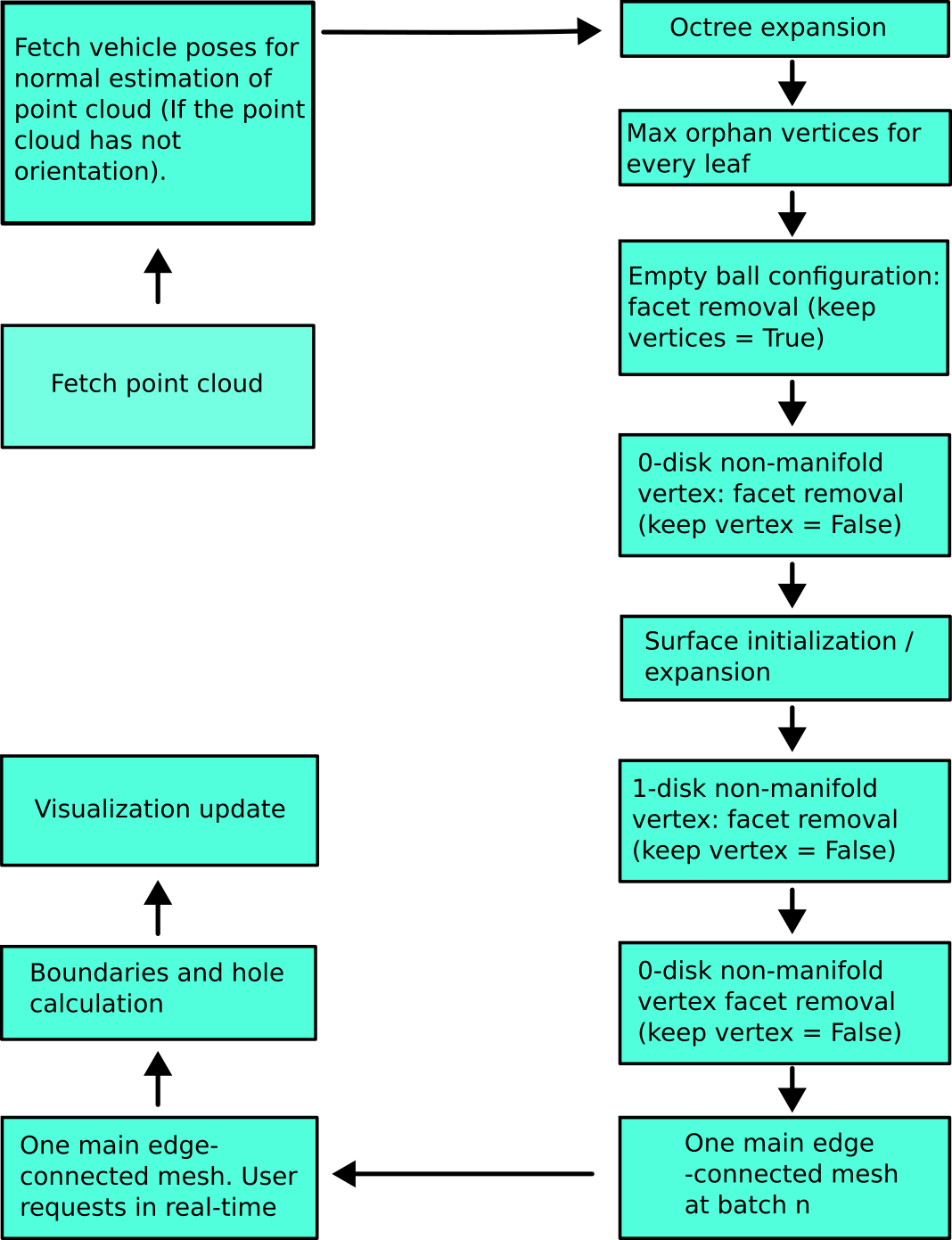}
    \caption{An overview of our IBPA pipeline. During facet removal, some approaches retain isolated vertices, whereas others remove vertices once they are no longer incident to any triangle. }
    \label{fig:system-overview}
\end{figure}

So far, we can incrementally extend an orientable manifold mesh (\cref{obj:1}, \cref{obj:2} and \cref{obj:3}).
In addition to the expandable octree, empty ball configuration enforcement, and non-manifold vertex removal, we now introduce several additional features to better align our method with real-world applications.
These enhancements are described in the following sections.
An overview of our IBPA pipeline is shown in \cref{fig:system-overview}.

\subsection{Normal Estimation}
The BPA algorithm requires the input point cloud to have associated normal vectors. However, in practice, not all robotic applications provide normal information for the point cloud.
To address this, we include a normal estimation module in our IBPA pipeline. For each vertex in the point cloud, we estimate the normal vector as pointing from the vertex toward the associated robot’s pose center.
As a result, our method also assumes access to the robot's pose in order to perform this normal estimation. However, this estimation can be disabled if the input point cloud already has orientation information.

\subsection{Max Orphan Vertices for Every Leaf Node in the Octree}
The reconstruction speed of IBPA is heavily influenced by the number of vertices within each neighboring leaf node. Since the size of a leaf is defined by the ball radius, which is a configurable parameter, this inherently sets the reconstruction resolution. If the user desires a higher resolution, the ball radius should be reduced, resulting in smaller leaf nodes.
Therefore, it is desirable to limit the number of orphan vertices within a single leaf node, as excessive orphan vertices increase the computational cost of BPA.
To manage computational load, we allow the user to define a maximum number of orphan vertices per leaf node. If the number of orphan vertices in a leaf exceeds this limit, we randomly select a subset of the orphan vertices equal to the specified maximum and remove the others.
%Note that this limit applies only to orphan vertices. A leaf node may still contain more total allowed orphan vertices, as not all vertices within it are orphan.

\subsection{Single Edge-connected Mesh and Hole Detection}
\citet{yip2024robust} classify the boundaries into primary boundaries, such as coastlines, and secondary boundaries, such as lakes or tide pools. Their method requires scanning the entire mesh, meaning that computation time increases proportionally with the number of facets. This design conflicts with the requirements for real-time processing.

Our goal is therefore to classify boundaries without the need to evaluate every facet in the existing mesh. The key idea is that if the mesh is already edge-connected, then re-evaluating the entire mesh is unnecessary. BPA inherently expands through the boundary-edges while preserving the edge-connected property of the mesh.

Boundaries are reconstructed from the set of boundary-edges. Since there are no non-manifold vertices, extracting boundary loops from this set is straightforward. We randomly select an edge from the boundary-edge set, follow the endpoint vertex to find the next connected edge, and repeat the process. No explicit search is required, as we use a pointer-based data structure, as illustrated in \cref{alg:class}. Once an edge is added to a boundary loop, it is removed from the boundary-edge set. This process continues until the set is empty and multiple boundary loops may be formed.

We define the longest boundary as the main boundary (e.g., coastlines), while all remaining boundaries are treated as holes (e.g., lakes). To implement this, we scan the entire mesh after processing batch~$n$ (a configurable parameter) and before batch~\(n+1\), retaining only the edge-connected mesh which has the longest boundary length and discarding any other edge-connected mesh.

\begin{figure}[htbp]
    \centering
    \begin{subfigure}[b]{0.48\textwidth}
        \centering
        \includegraphics[width=\textwidth]{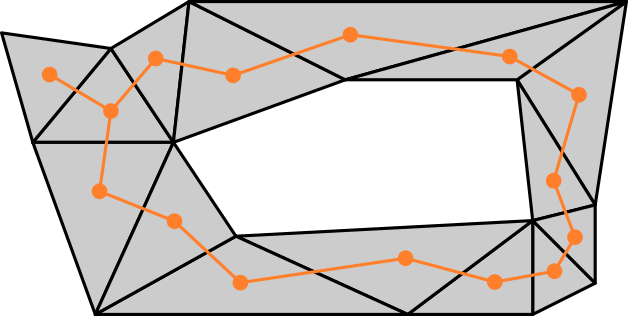}
        \caption{}
        \label{fig:graph-1}
    \end{subfigure}
    \hfill
    \begin{subfigure}[b]{0.48\textwidth}
        \centering
        \includegraphics[width=\textwidth]{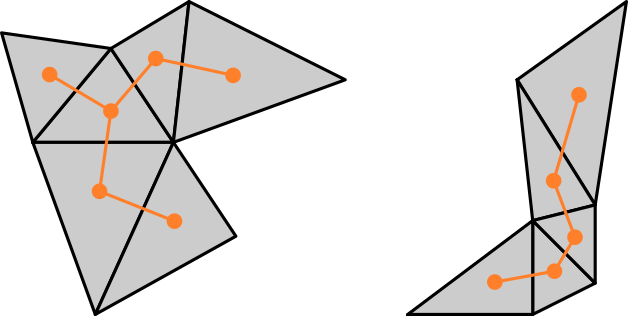}
        \caption{}
        \label{fig:graph-2}
    \end{subfigure}
    \caption{
    Illustration of a triangle mesh and its associated incentre graph. The orange vertices represent the incentres of the corresponding triangles. If two triangles are adjacent via a shared edge (but not merely a shared vertex), their incentres are connected by an edge in the graph. This construction yields a mathematical graph in which each vertex can have at most three edges.  
    (a) The graph $\mathcal{G}(\mathcal{V}, \mathcal{E})$, where $\mathcal{V}$ is the set of orange vertices and $\mathcal{E}$ is the set of orange edges, is connected—there exists a path between any pair of vertices.  
    (b) The graph is disconnected, as there exists at least one pair of vertices with no connecting path.
    }
    \label{fig:graph}
\end{figure}

Determining if a mesh is edge-connected is the same as determining whether its incentre graph (we refer to \cite{grimaldi2006discrete} for definitions of graphs and connectivity) is a connected graph. This relationship is depicted in \cref{fig:graph}. As shown in \cref{fig:graph-1}, an edge-connected mesh corresponds to a connected graph. In contrast, \cref{fig:graph-2} shows a mesh that lacks edge connectivity, along with its incentre graph, which is disconnected. To the best of our knowledge, evaluating all facets (or equivalently, all vertices in the incentre graph) is necessary to determine whether a mesh is edge-connected, or whether its incentre graph is connected.
%In contrast, Figure \cref{fig:graph-2} illustrates a mesh that lacks edge connectivity, with its dual graph that is not a connected graph. For the best of our knowledge, evaluating all facets (or vertices in the dual graph) is essential to conclude whether a mesh is edge-connected, or equivalently, whether its dual graph is connected.

%We can use the incentre of each triangle to define an inner vertex, and use connected edges between triangles to form edges between these inner vertices. This construction yields a graph, as shown in \cref{fig:graph}. We call this graph as incentre graph.
%Determining whether a mesh is edge-connected is equivalent to verifying that its incentre-based graph is connected (see \cite{grimaldi2006discrete} for definitions of graphs and connectivity). In Figure~\cref{fig:graph-1}, an edge-connected mesh corresponds to a connected graph, while in Figure~\cref{fig:graph-2}, a mesh that is not edge-connected produces a disconnected graph.
%To the best of our knowledge, verifying connectivity still requires visiting all facets. This can be done using standard graph traversal algorithms such as Depth-First Search (DFS) or Breadth-First Search (BFS). However, the traversal still necessitates visiting each facet. 
%We intentionally avoid using the term dual graph, as we currently lack a formal proof that the dual of the dual graph corresponds exactly to the original mesh.

\begin{figure}[htbp]
    \centering
    \begin{subfigure}[b]{0.48\textwidth}
        \centering
        \includegraphics[width=\textwidth]{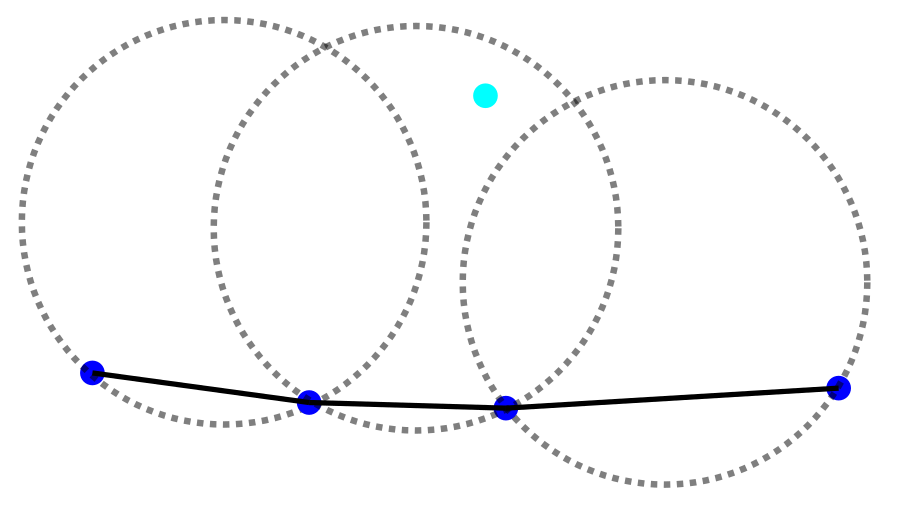}
        \caption{}
    \end{subfigure}
    \hfill
    \begin{subfigure}[b]{0.48\textwidth}
        \centering
        \includegraphics[width=\textwidth]{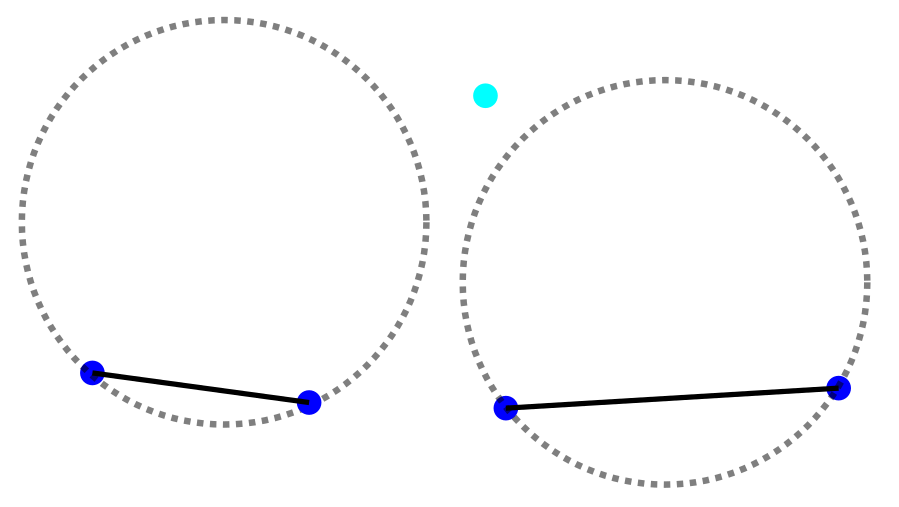}
        \caption{}
    \end{subfigure}
    \caption{An example illustrating how the removal of facets due to the empty ball configuration can split a single edge-connected mesh into two separate edge-connected components. In (a), a single edge-connected mesh contains an incoming vertex located inside the ball of an existing facet. As a result, our method removes the facet to preserve the empty ball configuration. However, this operation may unintentionally divide the original mesh into two edge-connected components, as shown in (b).}
    \label{fig:emptyConfigEdgeConnected}
\end{figure}

Although one might expect the mesh to remain a single edge-connected component, because BPA expands only through boundary-edges, this is not always the case. In rare scenarios, the removal of facets due to empty ball configuration violations can split an edge-connected mesh into two disconnected components (see \cref{fig:emptyConfigEdgeConnected}). This could lead to misclassification of boundaries.
Fortunately, such cases are rare and unlikely to be interpreted as missing information, since the user (a configurable parameter) can define a minimum hole length to qualify as an indication of missing information. Isolated fragments rarely exceed this threshold.
Although it is theoretically possible to run a full scan after each batch to preserve the whole mesh is a single edge-connected mesh, doing so would again violate real-time constraints. To mitigate this, we provide a manual override: The operator can press the shortcut key `t' allows to trigger a full connectivity check in cases where mesh fragmentation becomes visually apparent and problematic. During this process, fragmented mesh components, such as small isolated islands, will be detected and removed.

\section{Experimental results}
\label{sec:experiment}
We have implemented our method in C++, and we use Open3D \cite{zhou2018open3d} for real-time rendering and visualization. 
We evaluate our method in two operational modes: offline reconstruction and online reconstruction. 
%In this section, we evaluate our method in both offline and online settings.

\begin{figure}[htbp]
\centering
\includegraphics[width=0.65\textwidth]{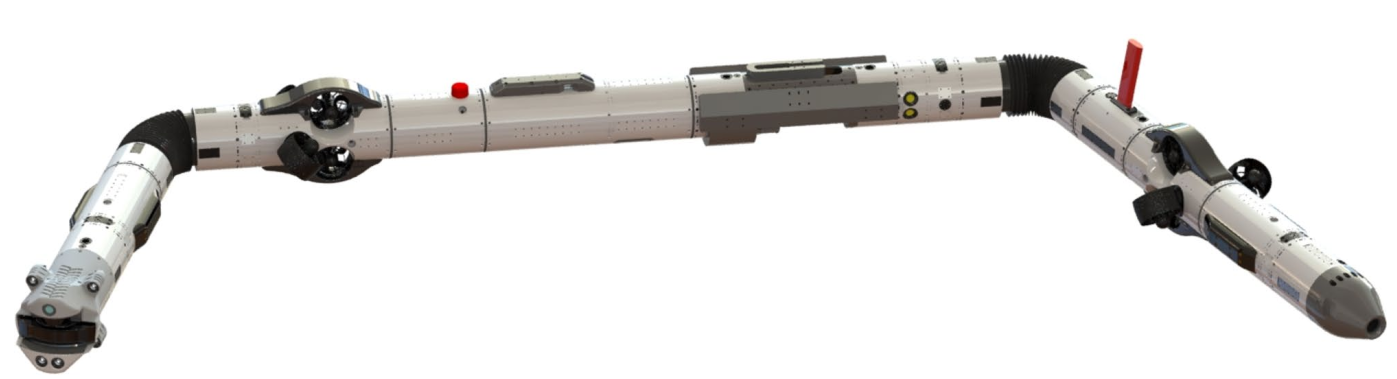}
\includegraphics[width=0.3\textwidth]{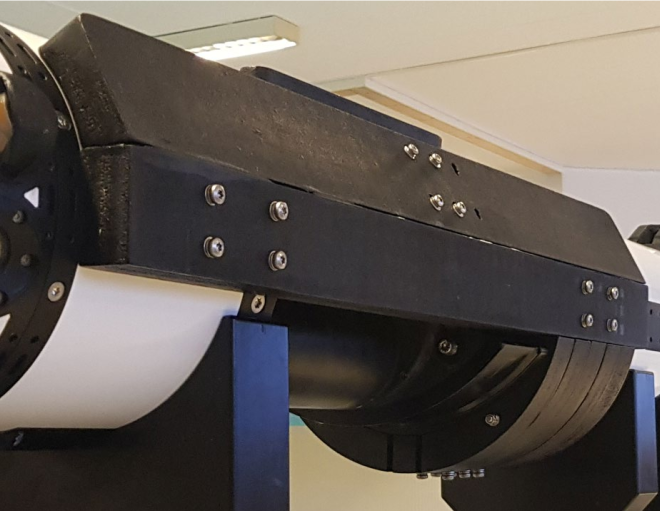}
\caption{Left: The Eelume robot (\cite{liljeback2017eelume}) with the MBES module. Right: A closer look of the MBES module.}
\label{fig:eelume-body}
\end{figure}

\subsection{Offline Experiments}
We introduce three real-world offline experiments. The respective point clouds were collected using an AUV, the Eelume underwater robot \cite{liljeback2017eelume} equipped with a multibeam echosounder (MBES), as shown in \cref{fig:eelume-body}.  
Kongsberg microPAP was used for navigation. The MBES operates at 
\(10\)Hz, and each scan produces around 250 points. Therefore, it generates approximately 2,500 points per second. %The MBES operates in \(10\)Hz and each scan produces around 250 points. Therefore, it is around 2500 points per second. 
The three datasets are \textit{Heinkel dataset}, \textit{Nyhavna dataset}, and \textit{Figaro dataset}.  
Both the Heinkel and Nyhavna datasets were collected in autonomous mode, whereas the Figaro dataset was acquired in remotely operated (ROV) mode using the same platform.

\begin{table}[htbp]
    \centering
    \begin{tabular}{|l|c|c|c|c|}
        \hline
        \textbf{Dataset} & \textbf{Mission time} & \textbf{Reconst. time} & \textbf{R} & \textbf{hole tolerance} \\
        \hline
        Heinkel & 886s & 22s & 1.0m & 3.16m \\
        \hline
        Nyhavna & 324s & 13s & 0.4m & 3.16m \\
        \hline
        Figaro & 547s & 7s & 1.0m & 3.16m \\
        \hline
    \end{tabular}
    \caption{Offline reconstruction performance comparison across different datasets.}
    \label{tab:offline_recon}
\end{table}
The Heinkel dataset contains sonar data of a submerged aircraft resting on the seafloor in the Trondheim fjord. 
The mission lasted 886 seconds, and our method completed the reconstruction in just $22$ seconds.
  
\begin{figure}[htbp]
    \centering
    \begin{subfigure}[b]{0.48\textwidth}
        \centering
        \includegraphics[width=\textwidth]{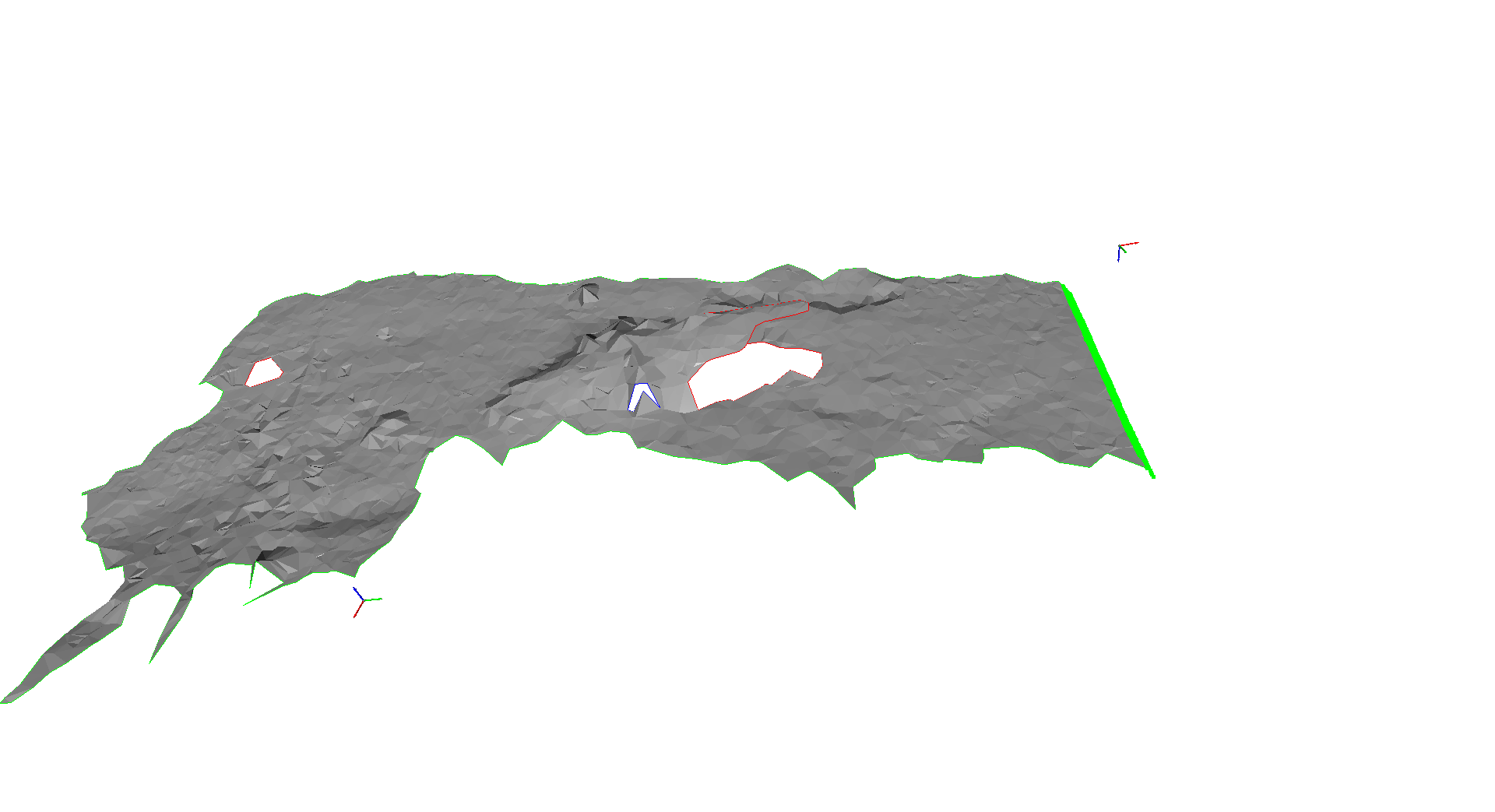}
    \end{subfigure}
    \hfill
    \begin{subfigure}[b]{0.48\textwidth}
        \centering
        \includegraphics[width=\textwidth]{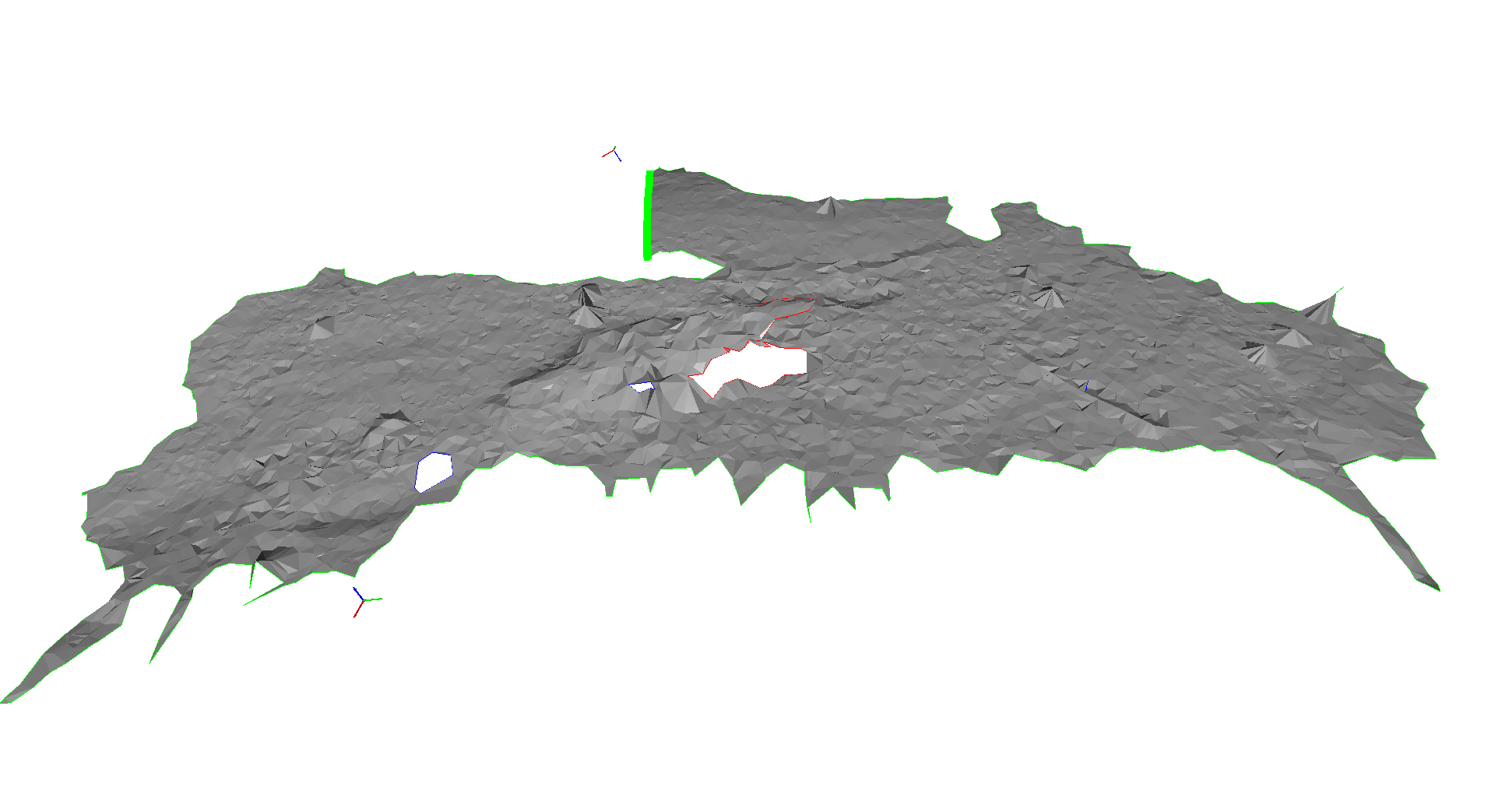}
    \end{subfigure}
    \caption{Incremental reconstrucion. Left: Reconstruction process at approximately $50\%$. Right: Reconstruction process at approximately $95\%$. }.
    \label{fig:incremental}
\end{figure}

\begin{figure}[htbp]
    \centering
    \begin{subfigure}[b]{0.48\textwidth}
        \centering
        \includegraphics[width=\textwidth]{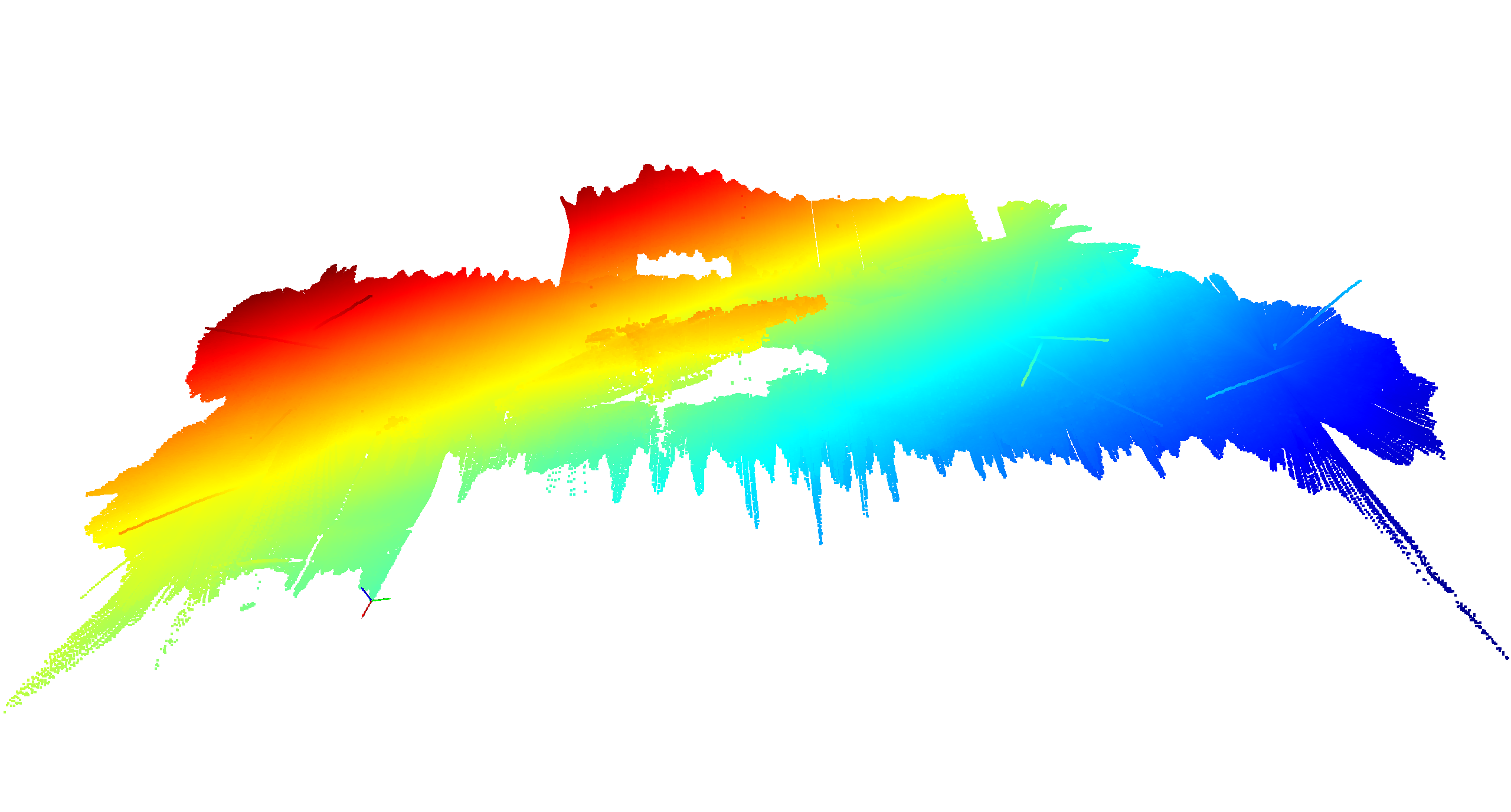}
        %\caption{Point cloud of Heinkel}
    \end{subfigure}
    \hfill
    \begin{subfigure}[b]{0.48\textwidth}
        \centering
        \includegraphics[width=\textwidth]{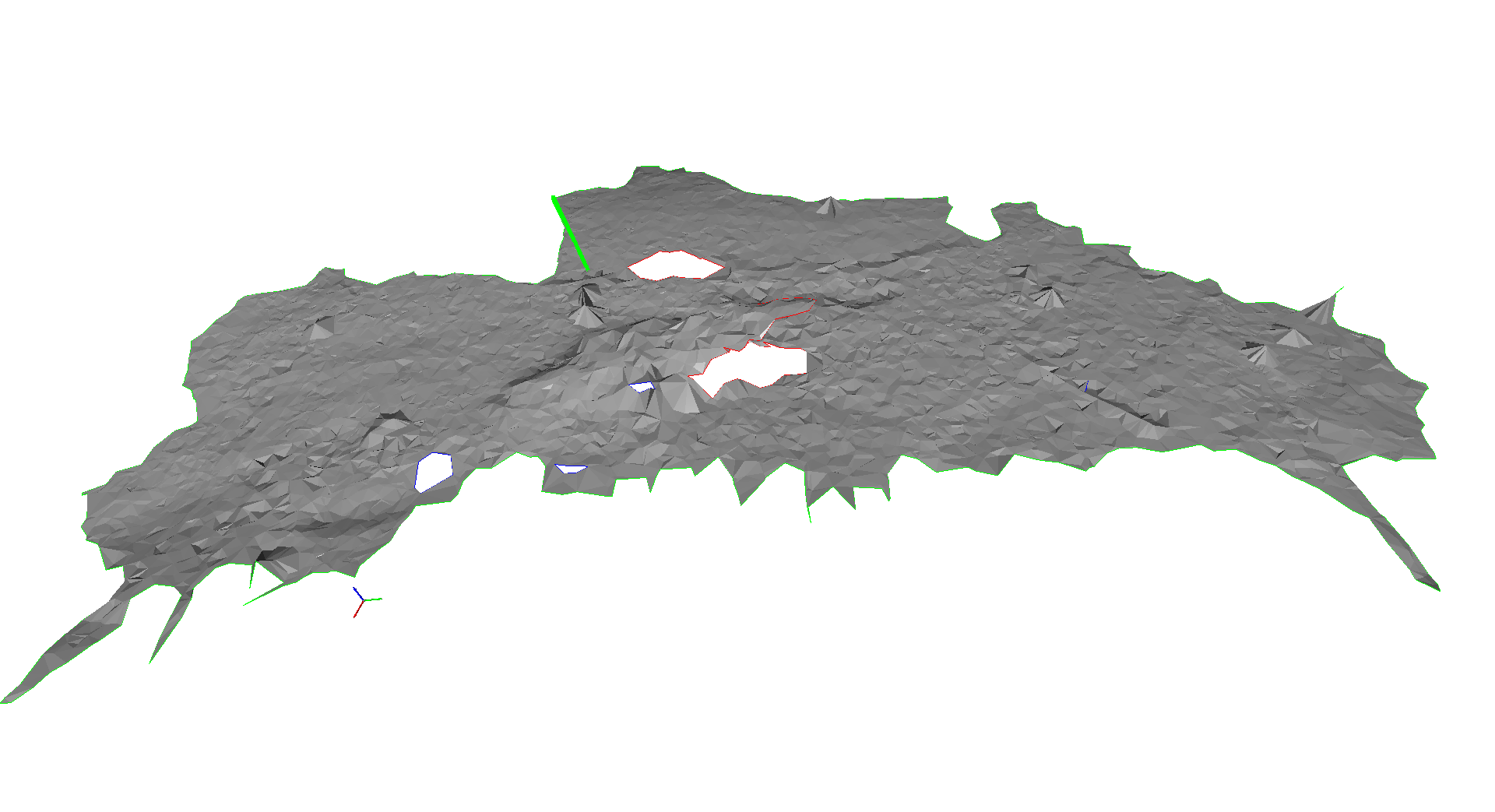}
        %\caption{Real-time surface reconstruction of Heinkel}
    \end{subfigure}
    \caption{Hole detection. Left: Point cloud of the Heinkel dataset. Visual inspection reveals two prominent holes.  
    Right: Length of the boundary of the hole longer than 3.16 meters are marked in red, while holes with length equal to or less than 3.16 meters are marked in blue.}
    \label{fig:Heinkel}
\end{figure}

\cref{fig:incremental} illustrates the incremental nature of our approach, showing the reconstruction at approximately $50\%$ and $95\%$ completion.
\cref{fig:Heinkel} shows the full point cloud of the dataset alongside the completed reconstructed mesh.  
As shown in \cref{fig:Heinkel} on the right, the AUV-collected data contain two prominent holes, both of which were successfully detected by our method during reconstruction.

\begin{figure}[htbp]
    \centering
    \begin{subfigure}[b]{0.95\textwidth}
        \centering
        \includegraphics[width=\textwidth]{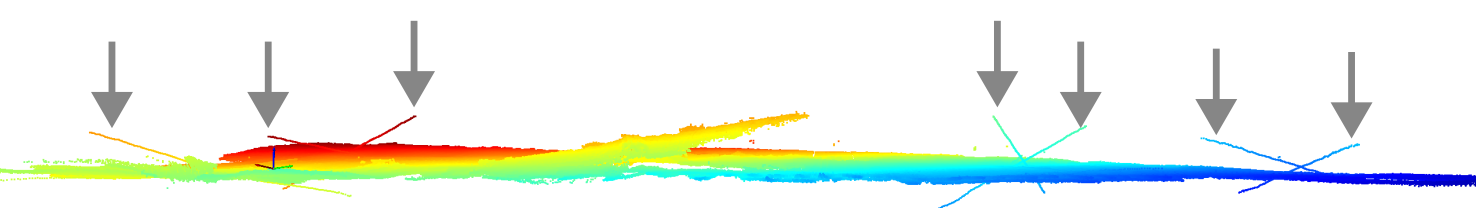}
        \label{fig:outlier-Heinkel-pcd}
    \end{subfigure}
    \vfill
    \begin{subfigure}[b]{0.95\textwidth}
        \centering
        \includegraphics[width=\textwidth]{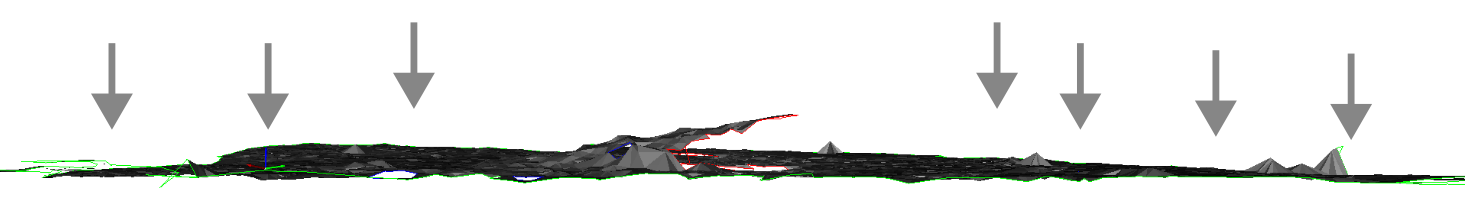}
        \label{fig:outlier-Heinkel-surf}
    \end{subfigure}
    \caption{Outlier handling. Top: Point cloud of the Heinkel dataset.  
    Bottom: Reconstructed surface of the Heinkel dataset.  
    Both figures are shown from the same viewport.  
    The gray arrow in the top figure highlights an outlier reading, caused by incorrect navigation georeferencing. The same arrows are shown in the bottom figure at the same position for reference.}
    \label{fig:heinkel_views}
\end{figure}

An interesting observation is the robustness of the  method against outlier readings.  
As shown in \cref{fig:heinkel_views}, there are seven outlier scan readings caused by incorrect georeferencing.  
However, our method effectively ignores most of these outliers or minimizes their impact.
%since the BPA-based method relies on local neighborhood
%the reconstruction ball cannot span large positional gaps caused by navigation errors.
%This ensures that only a minimal surface region is mistakenly reconstructed due to outlier points.

\begin{figure}[htbp]
    \centering
    \begin{subfigure}[b]{0.48\textwidth}
        \centering
        \includegraphics[width=\textwidth]{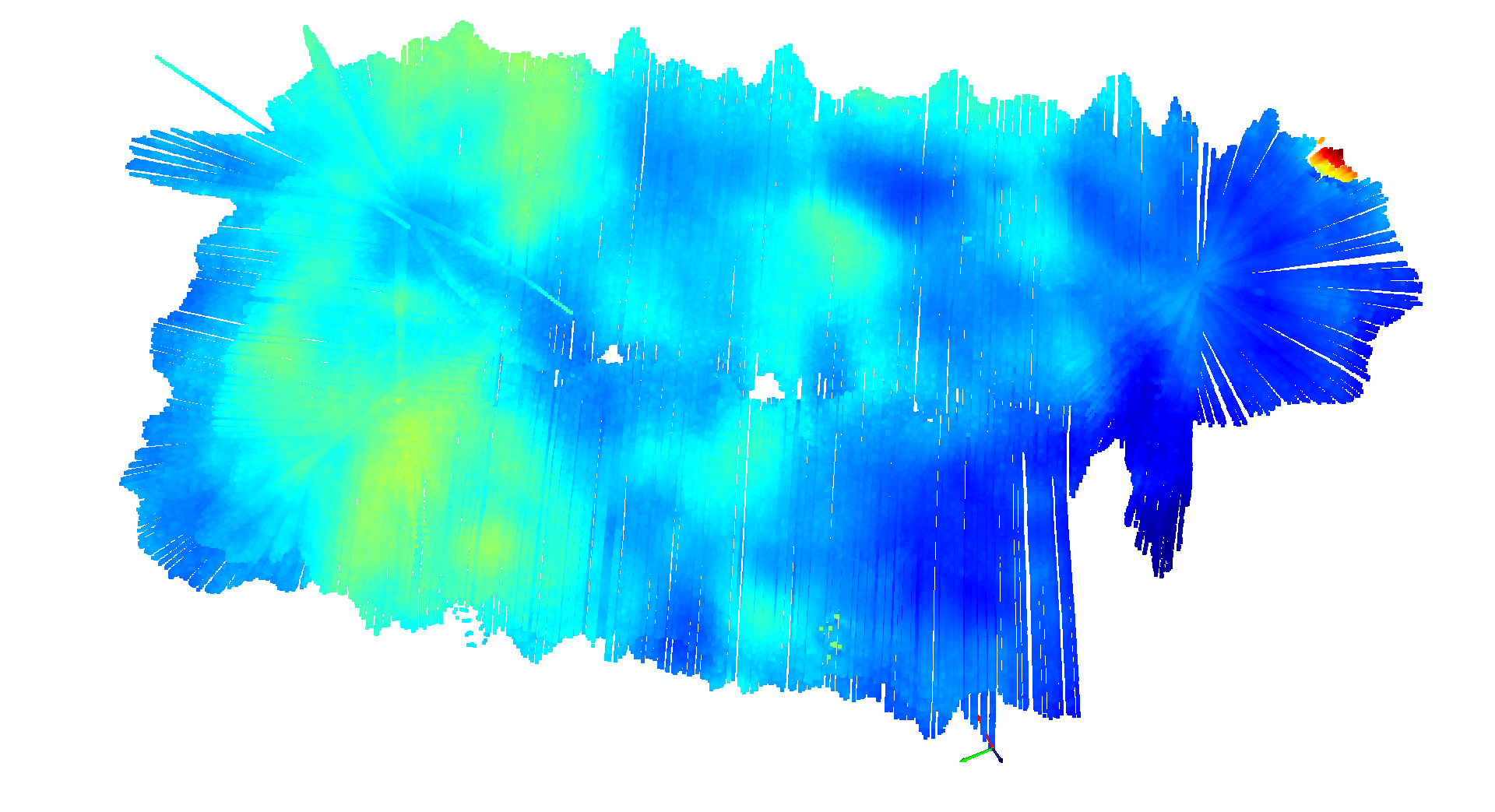}
        %\caption{Point cloud of Heinkel}
        \label{fig:nyhavna-pcd}
    \end{subfigure}
    \hfill
    \begin{subfigure}[b]{0.48\textwidth}
        \centering
        \includegraphics[width=\textwidth]{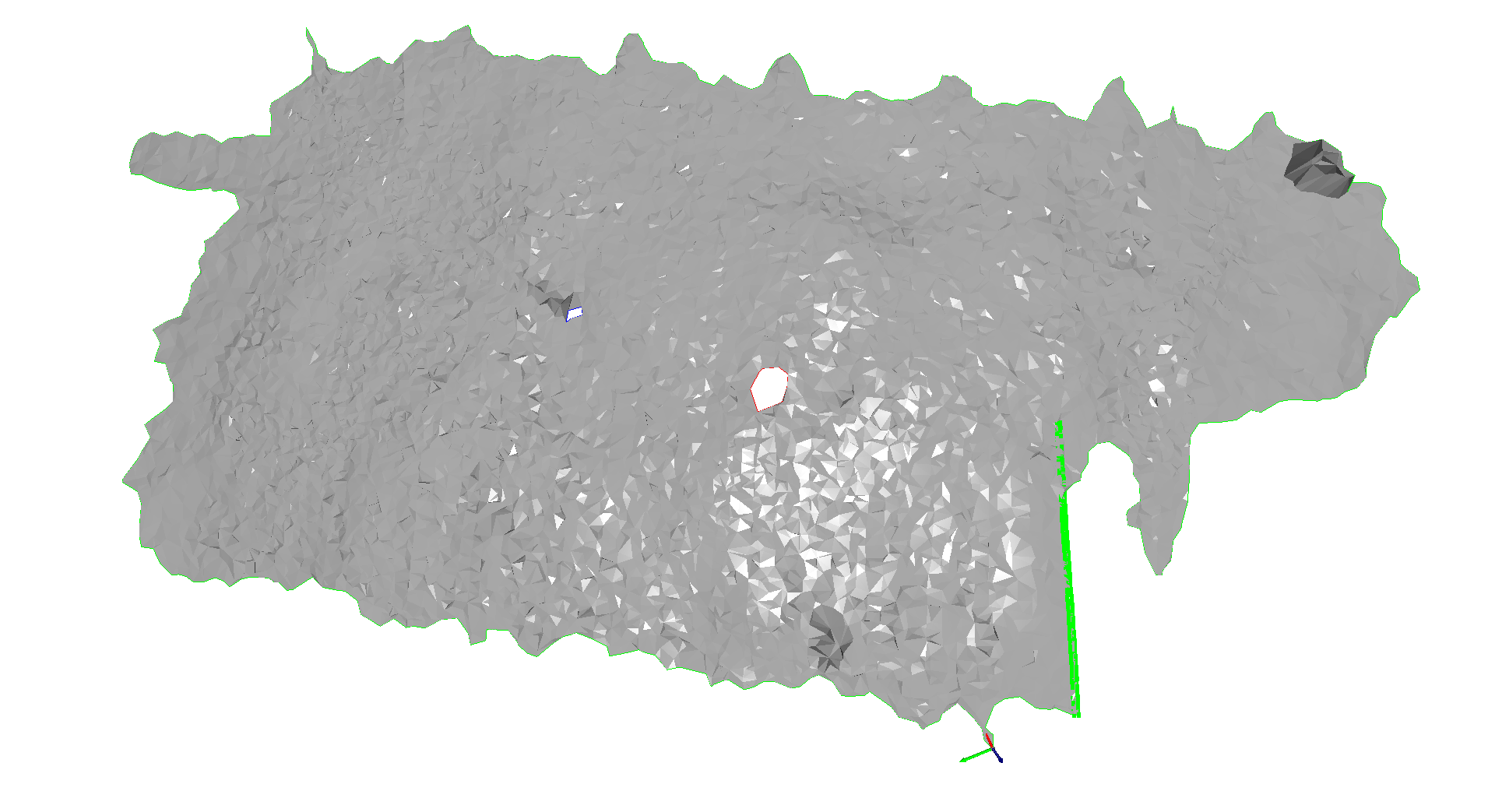}
        %\caption{Real-time surface reconstruction of Heinkel}
        \label{fig:nyhavna-surf}
    \end{subfigure}
    \caption{Hole detection. Left: Point cloud of the Nyhavna dataset.  
Right: Reconstructed surface of the Nyhavna dataset.  
One hole is detected in the point cloud.}
    \label{fig:nyhavna}
\end{figure}

For the Nyhavna dataset, the point cloud and reconstructed mesh are shown in \cref{fig:nyhavna}.  
The mission duration was $324$ seconds, and our method required $13$ seconds to reconstruct the mesh using a ball radius of $0.4$ meters.  
One hole was detected with a length exceeding the user-defined threshold and is marked in red.

\begin{figure}[htbp]
    \centering
    \begin{subfigure}[b]{0.48\textwidth}
        \centering
        \includegraphics[width=\textwidth]{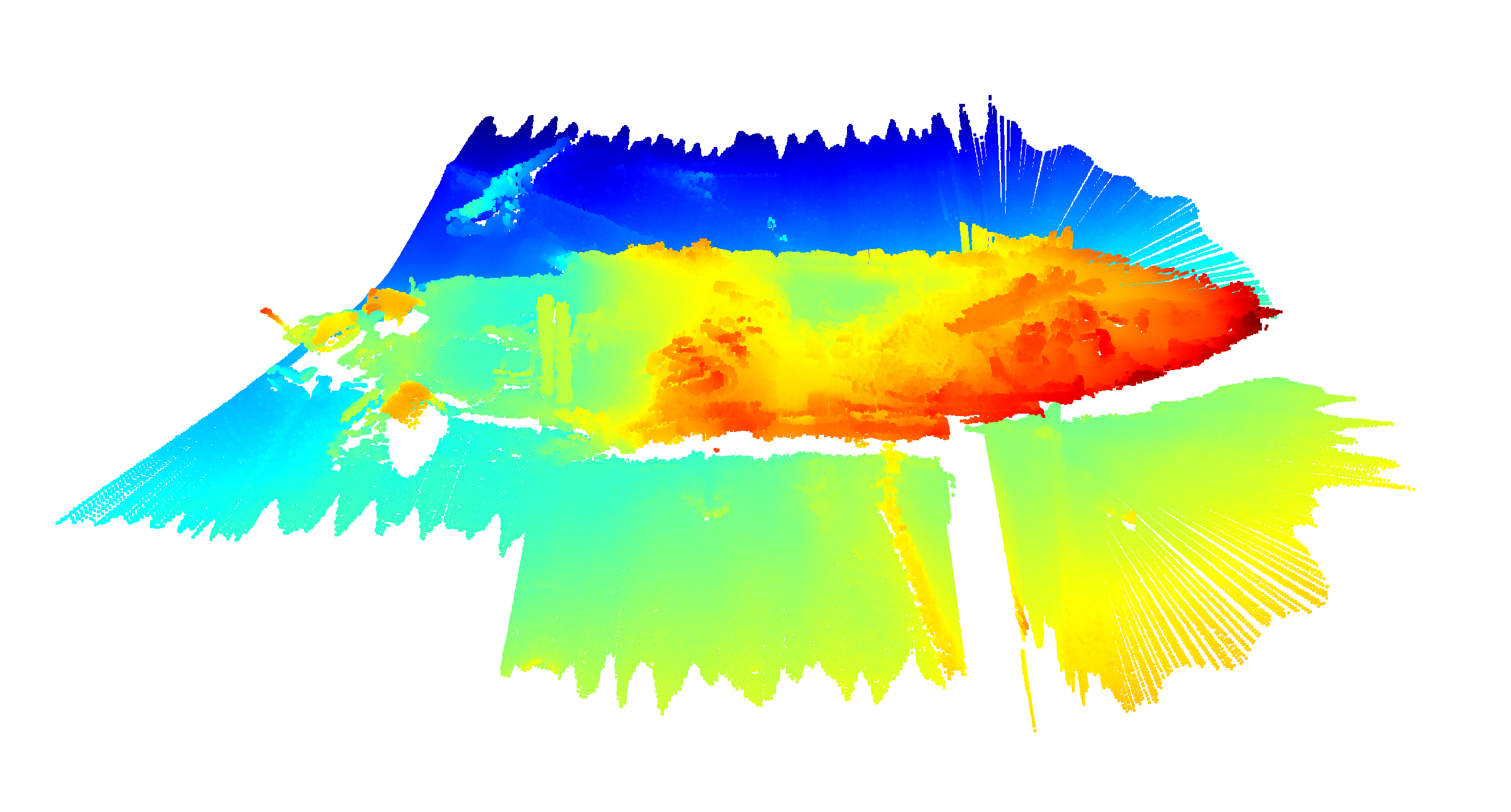}
        %\caption{Point cloud of Heinkel}
        \label{fig:figaro-pcd}
    \end{subfigure}
    \hfill
    \begin{subfigure}[b]{0.48\textwidth}
        \centering
        \includegraphics[width=\textwidth]{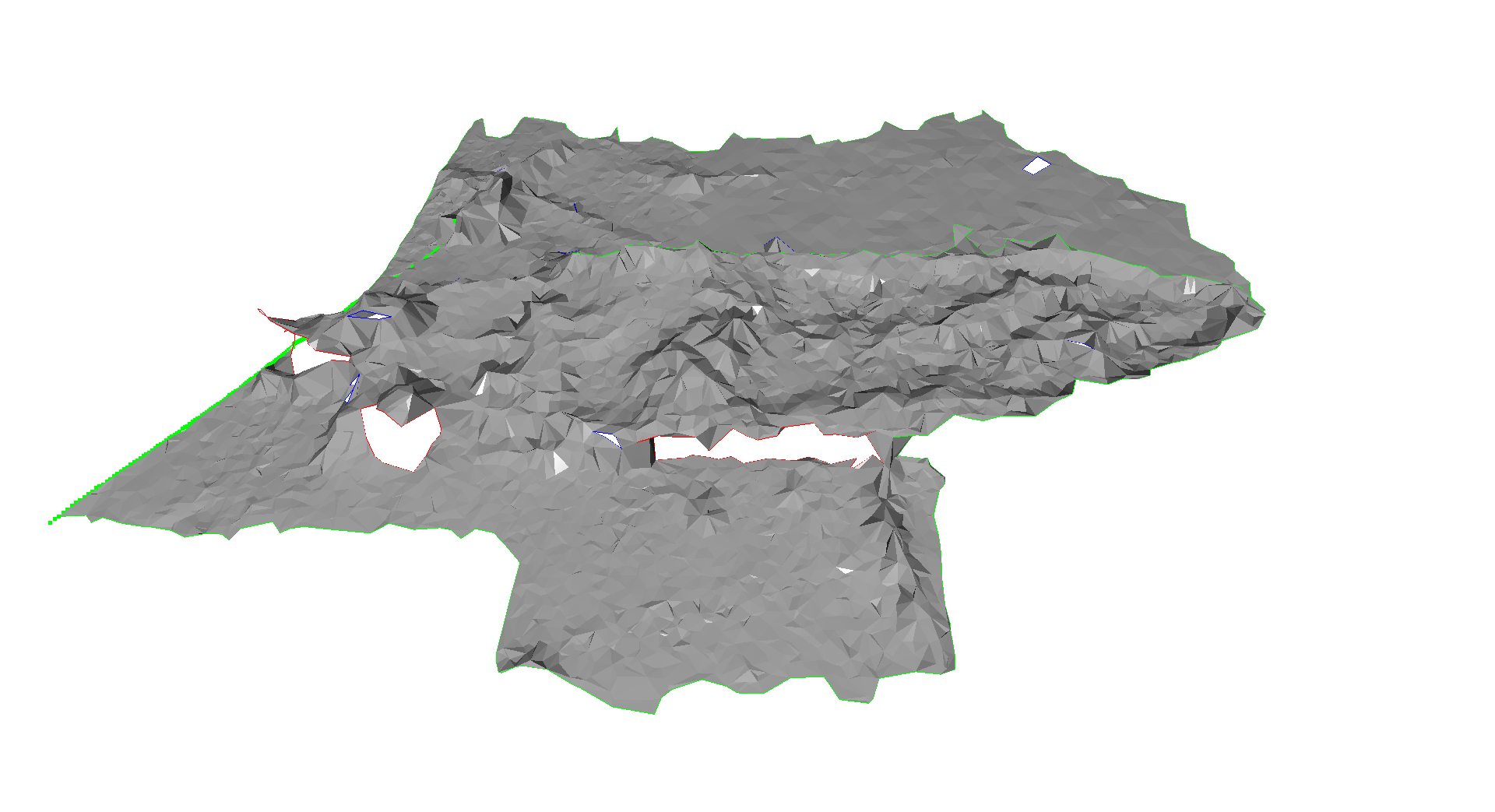}
        %\caption{Real-time surface reconstruction of Heinkel}
        \label{fig:figaro-surf}
    \end{subfigure}
    \caption{Consistency check. Left: Point cloud of Figaro dataset. Right: Reconstructed surface. A translational displacement is visible in the bottom-right corner of the point cloud. This demonstrates that the BPA-based method inherently performs local consistency checks, rejecting disconnected regions unless the radius is sufficiently large to span the translation error.}
    \label{fig:figaro}
\end{figure}

\cref{fig:figaro} shows the point cloud of the Figaro dataset and the reconstructed surface.  
The mission time was $547$ seconds, and our method required only $7$ seconds to reconstruct the mesh using a radius of $1$ meter.  
Three large holes were detected, using a tolerance \(3.16\) meters.  
Note that the bottom-right portion of the point cloud exhibits a translational shift to the right.  
Our method correctly identifies this region as an outlier, since the BPA ball is not large enough to connect it to the main structure.

\subsection{Online Experiments}
\begin{figure}[htbp]
  \begin{subfigure}{0.9\textwidth}
    \centering
    \includegraphics[width=\linewidth]{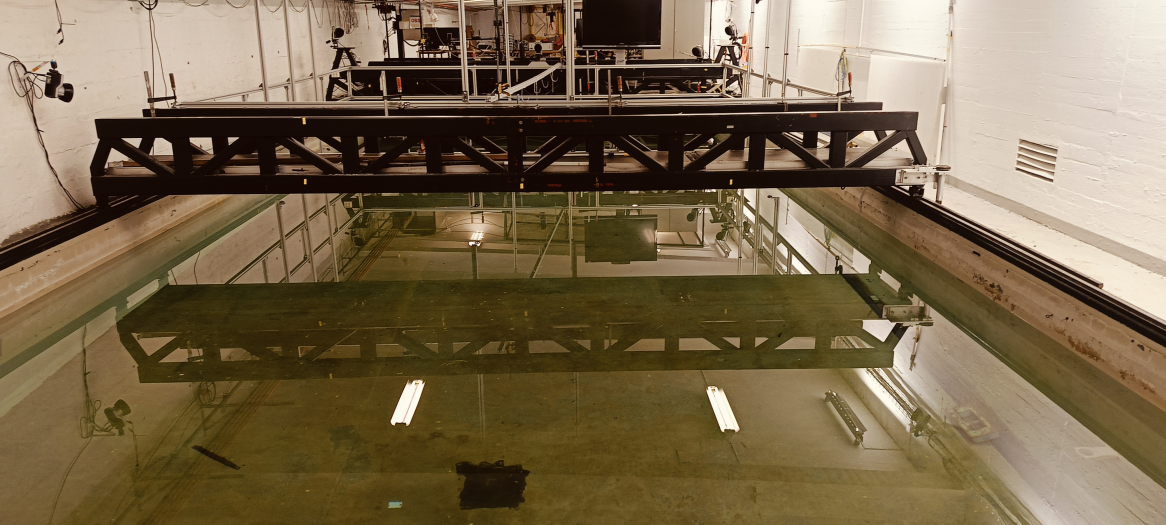}
    \caption{}
    \label{fig:online-indoor-1}
  \end{subfigure}
    \vspace{0.5em}  % Optional vertical space between rows
  \centering
  \begin{subfigure}{0.49\textwidth}
    \includegraphics[width=\linewidth]{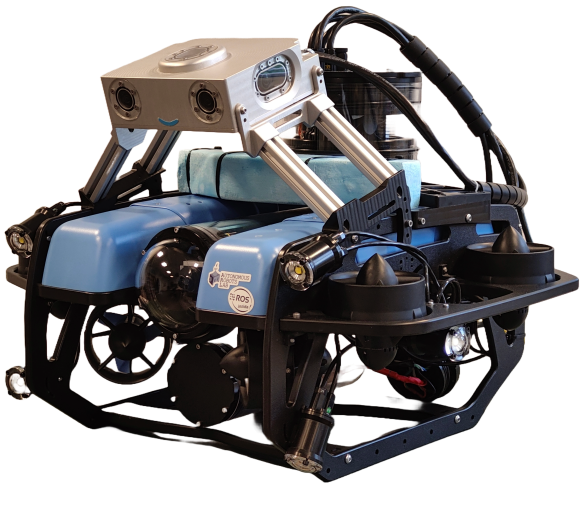}
    \caption{}
    \label{fig:online-indoor-3}
  \end{subfigure}
  \begin{subfigure}{0.49\textwidth}
    \includegraphics[width=\linewidth]{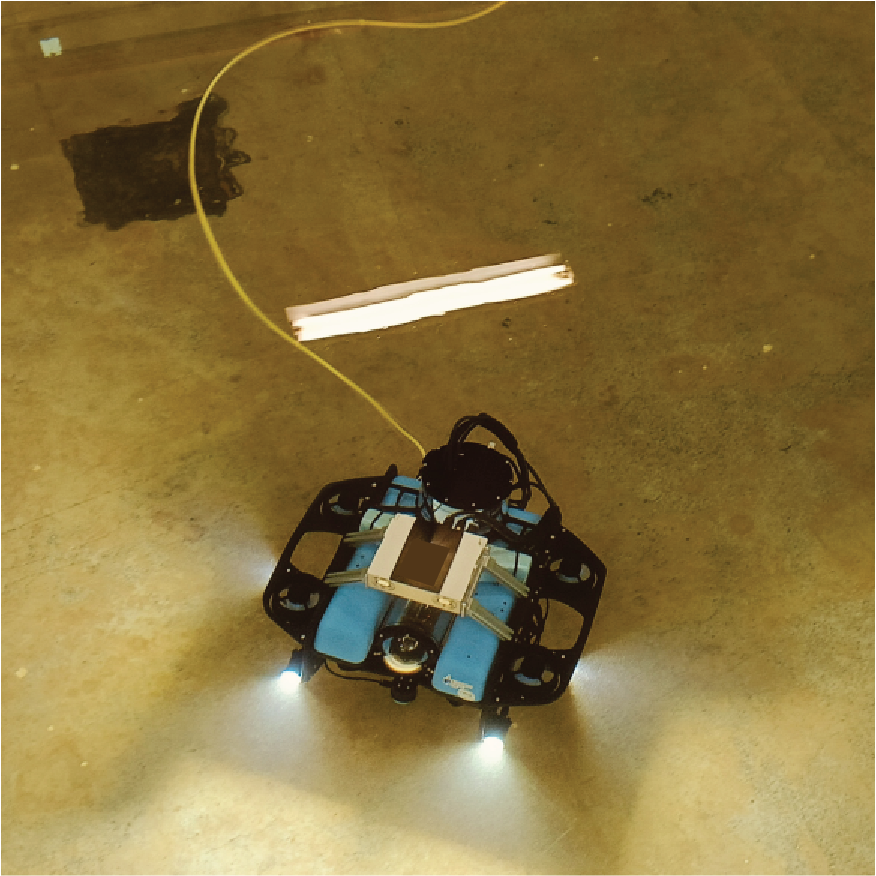}
    \caption{}
    \label{fig:online-indoor-2}
  \end{subfigure}
  \caption{(a): The Marine Cybernetics Laboratory (MC-Lab) at NTNU.
            (b): The Ariel underwater robot. 
            (c): The robot submerged in water.
            }
  \label{fig:online-indoor}
\end{figure}

\begin{figure}
    \centering
    \includegraphics[width=0.9\linewidth]{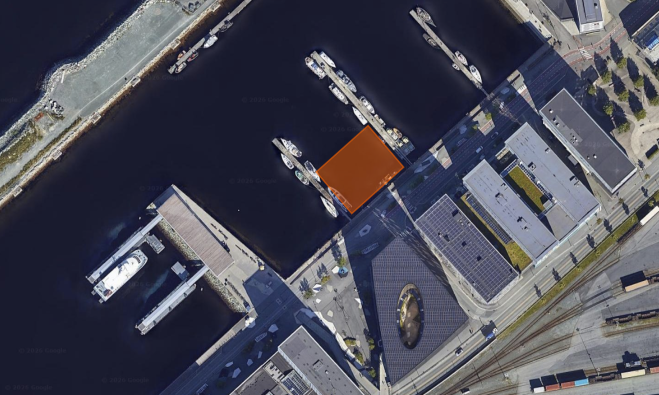}
    \caption{Trondheim Harbour experimental site. The mapping area is marked in orange. Imagery \textcopyright{}2026 Airbus, CNES / Airbus, Maxar Technologies; map data \textcopyright{}2026 Google}
    \label{fig:trondheimHarbour}
\end{figure}

We present two online experiments conducted with an underwater robot equipped with a vision-based multi-camera navigation stack, as described in \cite{ReAqROVIO}, alongside velocity aiding from a learning based proprioceptive method \cite{DeepVL}. The first experiment was carried out in a controlled environment at the Marine Cybernetics Laboratory at NTNU (see \cref{fig:online-indoor-1}). \cref{fig:online-indoor} shows the experimental indoor setup and the underwater robot. The second took place in the outdoor setting of the Trondheim Harbour (see \cref{fig:trondheimHarbour}). The robot relies on vision-based state estimation and perception, using visual-inertial odometry (VIO) to estimate its relative 6-degree-of-freedom (DoF) pose and a learning-based stereo depth estimation approach for perception and mapping. 

%We demonstrate two online experiments with an underwater robot using a multi-camera vision-based navigation stack as described in \cite{ReAqROVIO}. The first experiment took place in a controlled laboratory environment in the Marine cybernetics laboratory at NTNU, see Figure \cref{fig:online-indoor-1}. The second experiment was carried out in the outdoor setting of the Trondheim fjord, see figure \cref{fig:trondheimFjord}. The robot functions on a vision-based state estimation and perception, it uses visual-inertial odometry (VIO) to obtain relative 6 degree-of-freedom (dof) pose and a learning-based stereo camera depth estimation for perception and mapping. Figure \cref{fig:online-indoor} illustrates the experimental indoor setup and the underwater robot.

% In the online experiments, we present two experiments with the robot, as detailed in \cite{ReAqROVIO}. One experiment took place in a controlled laboratory environment in Marine cybernetics laboratory at NTNU, see figure \cref{fig:online-indoor-1}.
% The second experiment was carried out in the outdoor setting of the Trondheim fjord, see figure \cref{fig:trondheimFjord}. The robot does not have an acoustic device and instead uses an IMU and visual odometry to approximate its relative position in relation to the surrounding point cloud. Figure \cref{fig:online-indoor} illustrates the experimental indoor setup and the underwater robot.

\begin{figure}[htbp]
  \centering
  \begin{subfigure}{0.49\textwidth}
    \includegraphics[width=\linewidth]{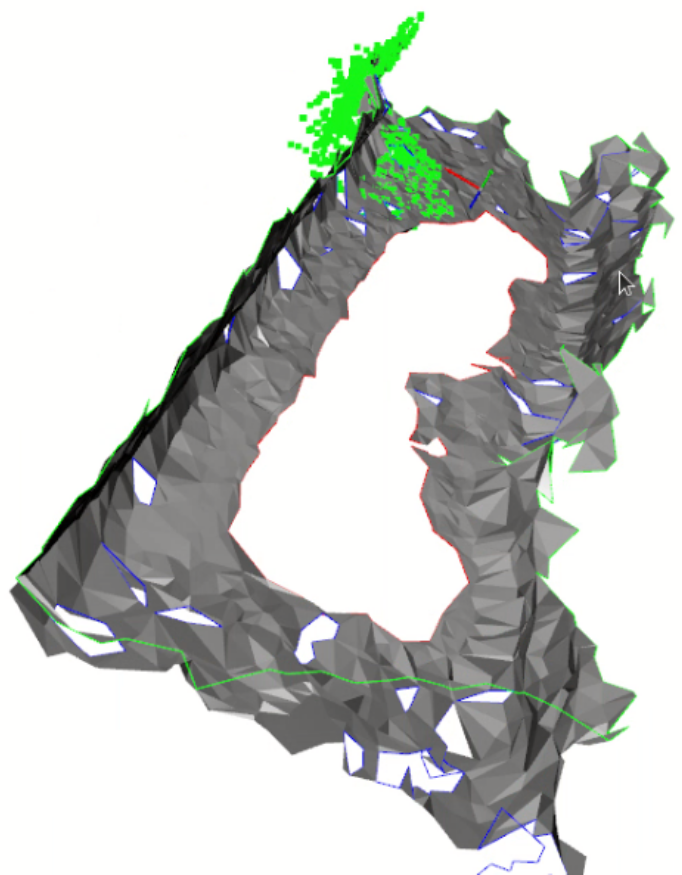}
    \caption{}
    \label{fig:online-indoor-hole}
  \end{subfigure}
  \begin{subfigure}{0.49\textwidth}
    \includegraphics[width=\linewidth]{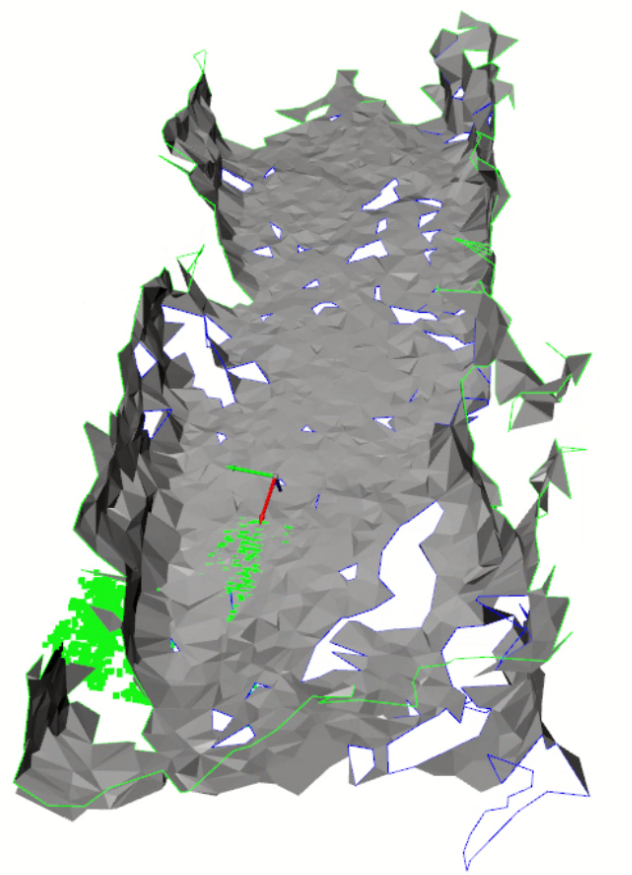}
    \caption{}
    \label{fig:online-indoor-fill}
  \end{subfigure}
  \caption{Results of our IBPA method with missing information detection are shown. The back facets are configured to be transparent. (a): A hole with boundary length exceeding $3$ meters in length was detected, shown as red boundary. (b): Utilizing the information from (a), the hole was filled.
}
  \label{fig:online-indoor-exp}
\end{figure}

In the controlled indoor experiment, the robot was commanded to follow the tank wall in a circular trajectory to evaluate our method’s ability to detect missing information. After detecting the hole, the operator performed an additional scan to acquire data over the missing region, which was then automatically integrated by our method (\cref{fig:online-indoor-fill}). The radius of the ball was set to 0.5 meters, and a gap (or``hole") of 3 meters was intentionally introduced. \cref{fig:online-indoor-hole} shows the result, where a red boundary highlights the detected gap, measuring more than 3 meters in length. After identifying this missing information, 
the robot was manually steered back to gather additional data to fill the hole, as shown in \cref{fig:online-indoor-fill}.
%In the controlled online indoor experiment, we control the robot to follow the tank's wall to form a large circle, testing our method's ability to detect missing information. We configure the ball's radius to be $0.5$ meters and set the hole length to $3$ meters. 
%Figure \cref{fig:online-indoor-hole} shows the result. A red boundary is depicted at the center, the length of the hole is more than $3$ meters. Upon identifying this missing information, we acquired more information to fill the hole, as depicted in Figure \cref{fig:online-indoor-fill}.

\begin{figure}[htbp]
  \centering
  \begin{subfigure}{0.49\textwidth}
    \includegraphics[width=\linewidth]{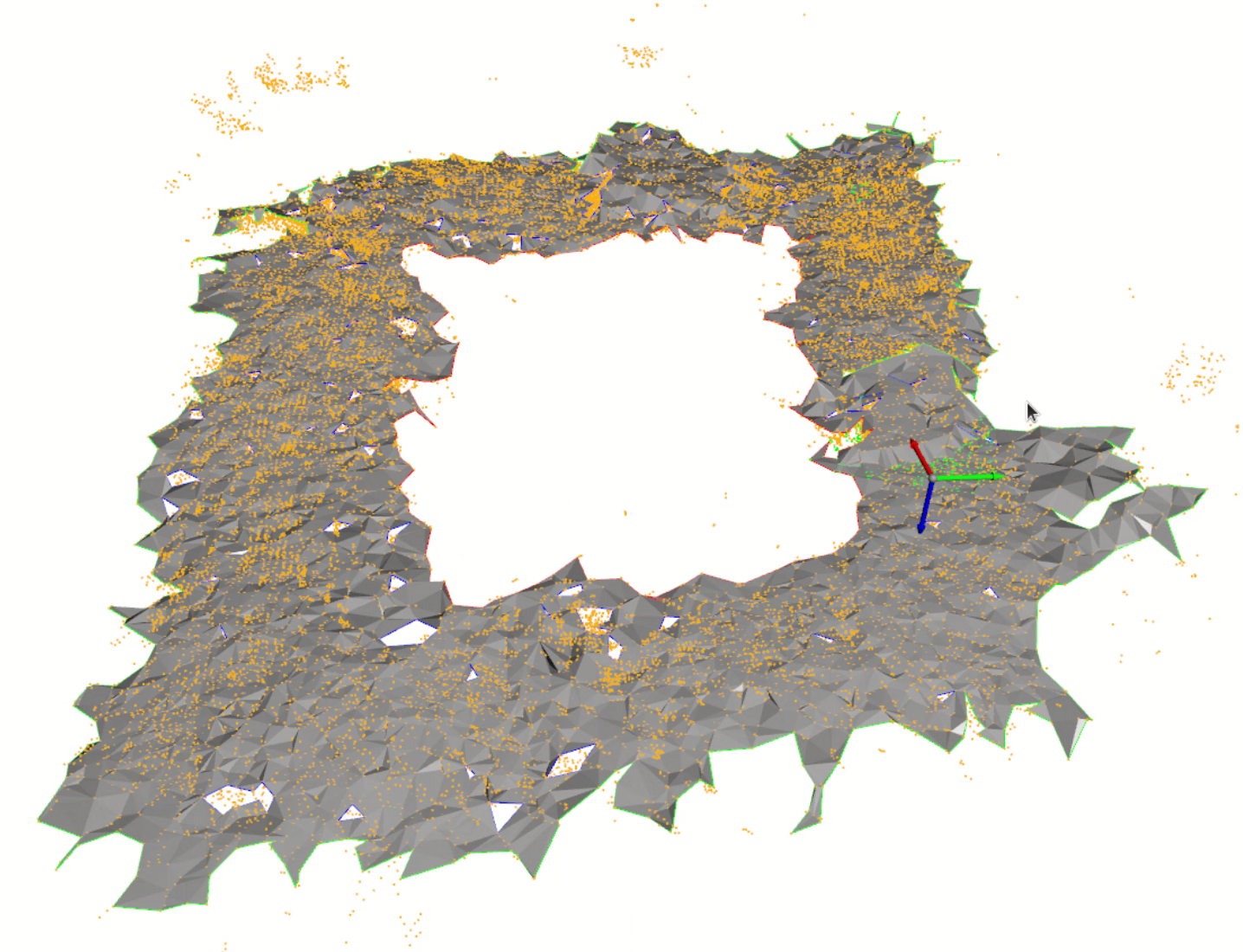}
    \caption{}
    \label{fig:online-outdoor-hole1}
  \end{subfigure}
  \begin{subfigure}{0.49\textwidth}
    \includegraphics[width=\linewidth]{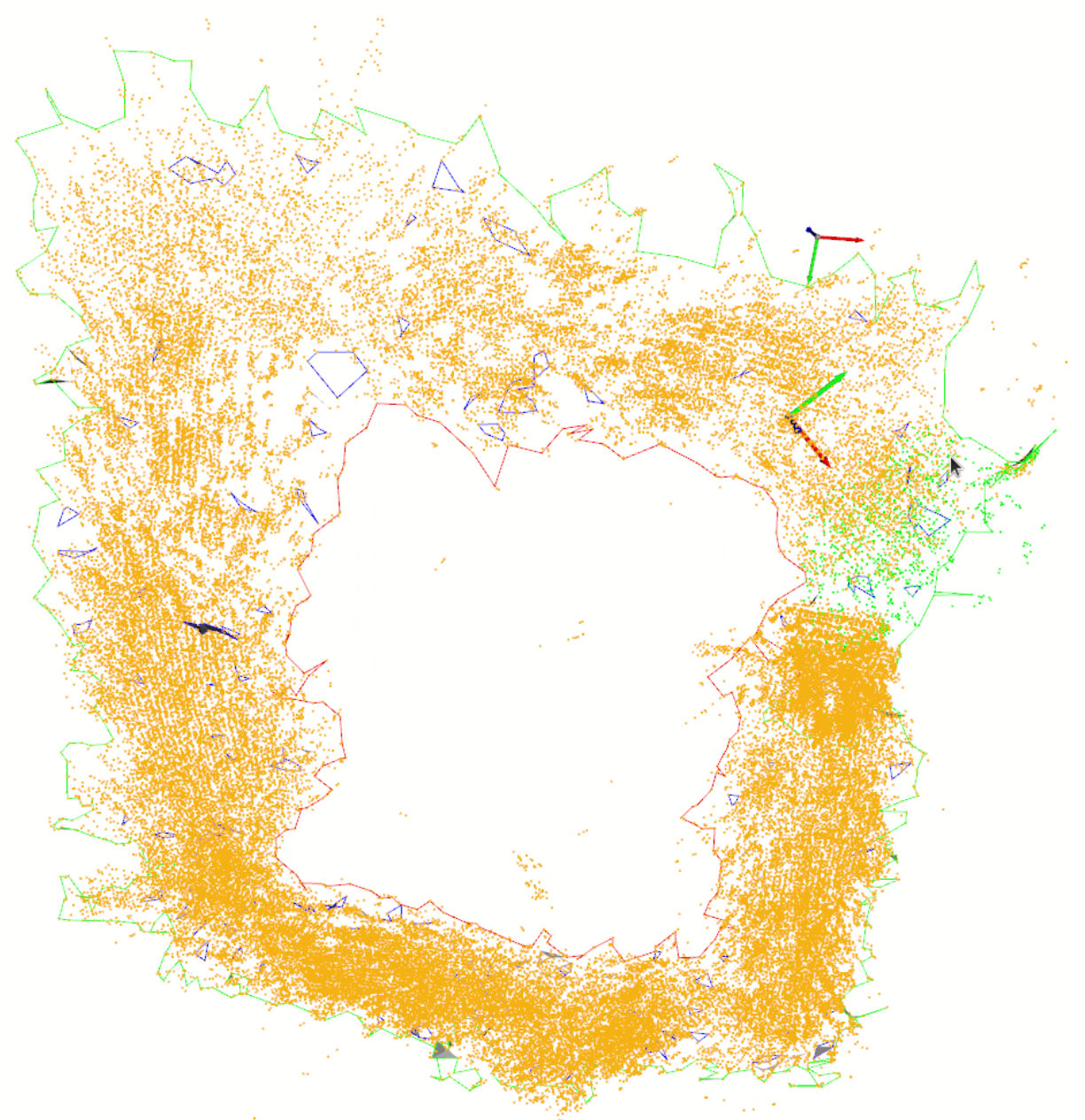}
    \caption{}
    \label{fig:online-outdoor-hole2}
  \end{subfigure}
  \caption{Hole detection in an online outdoor experiment. The yellow point cloud represents all points previously received from the robot by our algorithm. (a) The detected hole measures more than 3 meters in boundary length. (b) View of the reconstructed surface from below. The mesh is rendered with back-face transparency enabled to better visualize the holes and validate the detection.}
  \label{fig:online-outdoor-hole}
\end{figure}

\begin{figure}
  \begin{subfigure}{0.95\textwidth}
    \centering
    \includegraphics[width=\linewidth]{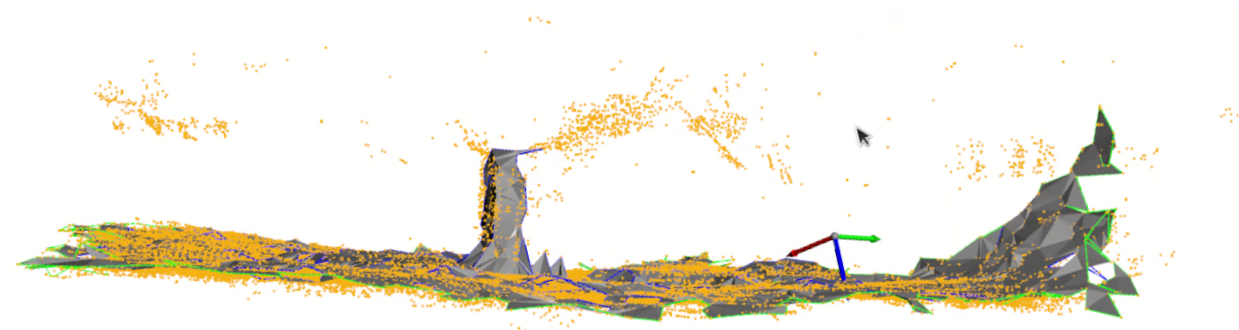}
    \caption{}
    \label{fig:online-indoor-5}
  \end{subfigure}
  \vspace{0.5em}
  \begin{subfigure}{0.95\textwidth}
    \centering
    \includegraphics[width=\linewidth]{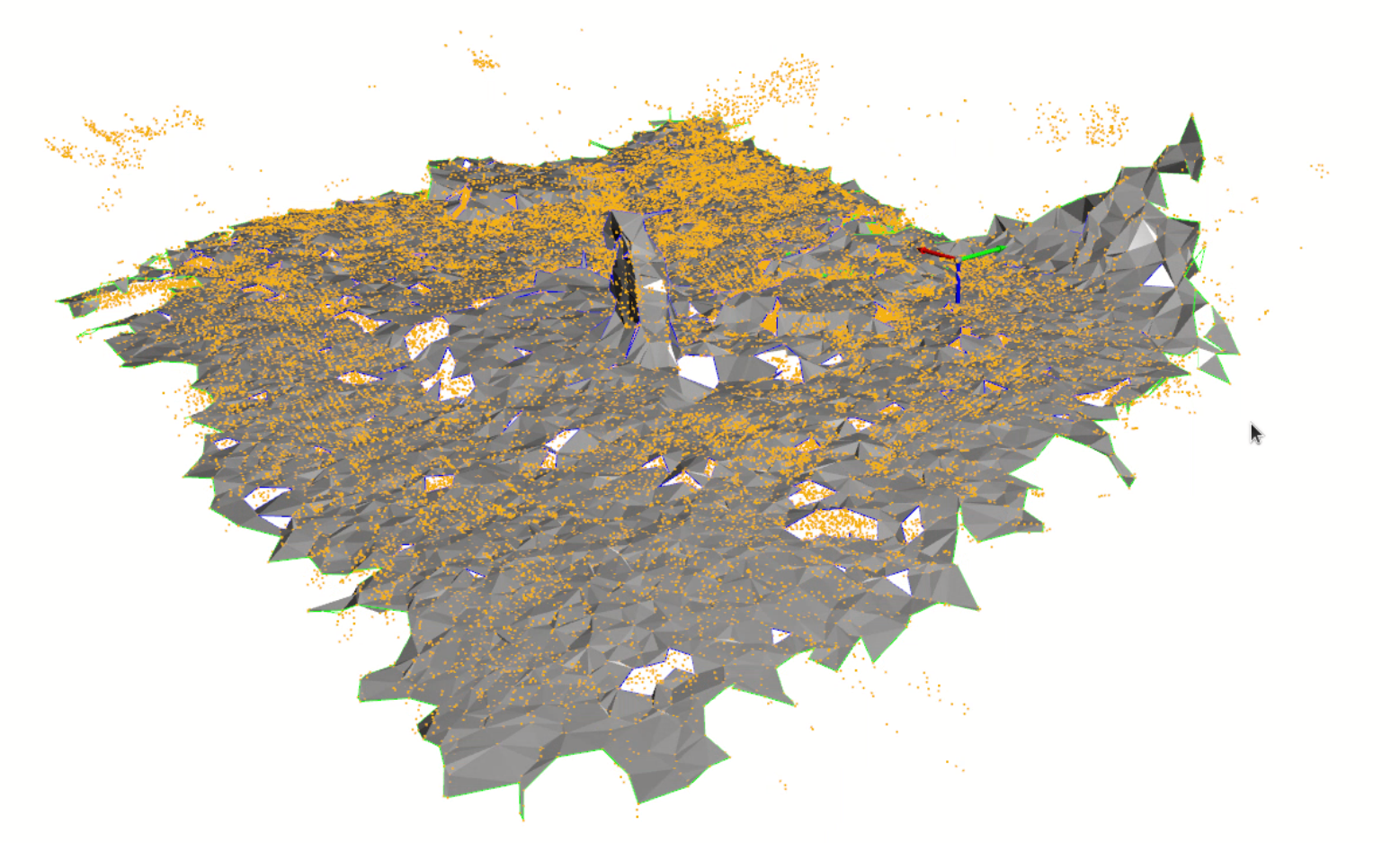}
    \caption{}
    \label{fig:online-indoor-4}
  \end{subfigure}
  \caption{The surface reconstruction was successfully completed with minimal missing information, as all hole boundaries are shorter than 3 meters (blue boundaries only). In (a), outlier points are visible near the water surface, which our method effectively removes. A central pillar appears to contact the water surface due to an actual rod-like structure extending from the seabed nearly to the water column. The right side of (a) shows the harbor wall. In (b), the final surface reconstruction is presented with no significant missing data (no red boundaries).}
    \label{fig:online-outdoor-finish}
\end{figure}

In the outdoor experiment, we remotely controlled the robot to map the area of Trondheim Harbour marked in \cref{fig:trondheimHarbour}, while our method reconstructed the surface online. \cref{fig:online-outdoor-hole} shows an intermediate state of the reconstruction, where the yellow point cloud represents all points received from the robot so far. In \cref{fig:online-outdoor-hole1}, a hole with a boundary length of more than 3 meters was detected and highlighted with a red boundary. To verify this detection, \cref{fig:online-outdoor-hole2} shows the reconstructed surface from below with back-face transparency enabled, which makes the holes in the mesh clearly visible. After additional data had been collected over this region, the reconstruction was completed with minimal missing information, as shown in \cref{fig:online-outdoor-finish}: all remaining hole boundaries are shorter than 3 meters and are therefore marked in blue only. \cref{fig:online-indoor-5} also contains outlier points near the water surface, which our method effectively removes, and shows the harbor wall on the right-hand side. The central pillar that appears to reach the water surface is not a reconstruction artifact; it corresponds to an actual rod-like structure extending from the seabed nearly to the surface. The final surface reconstruction in \cref{fig:online-indoor-4} contains no significant missing data, as indicated by the absence of red boundaries.

In both online experiments (Harbour + indoor), we deliberately maneuver the robot so that missing triangles occur in the center of the geometric experimental space, in order to demonstrate that our method can detect holes. Nevertheless, our method will successfully detect any hole once its length exceeds the user-specified threshold, irrespective of the hole's irregular shape.
\clearpage
\section{Conclusion} \label{sec:concl}
We have extended the original Ball Pivoting Algorithm (BPA) to operate in an incremental manner, which we refer to as Incremental BPA (IBPA). This includes enabling the octree to expand dynamically as needed and maintaining the empty ball configuration by removing any violating facets, without deleting vertices.

To preserve the manifold structure, we scan all boundary-edges to detect non-manifold vertices. For $1$-disk and $0$-disk cases, any vertex violating manifold conditions triggers removal of the associated facets. This remains efficient, as the number of boundary-edges grows slowly relative to the full mesh.

We repeat this process until all vertices in the current batch are manifold. To maintain orientability, we also verify consistent orientation among neighboring facets during surface reconstruction and discard any that violate this condition.

To limit reconstruction complexity, we limit the number of orphan vertices per leaf node to a user-defined value $n$. The physical size of a leaf is determined by the ball radius, which also sets the reconstruction resolution.

We presented three offline experimental datasets collected by the Eelume robot equipped with a multibeam echosounder. These experiments demonstrate that our method is capable of handling outlier samples to a good extent. Moreover, it successfully identifies holes whose boundaries exceed a user-defined tolerance length and highlights them with a red boundary to inform the user. In addition, we presented two online experiments to demonstrate the applicability of our method in real-time and outdoor settings. These results confirm that IBPA is suitable for both batch and stream reconstruction pipelines in practical underwater robotics scenarios.

We have open-sourced our code with full ROS integration, enabling easy deployment on custom robotic systems.

Future work will focus on:  
(1) improving efficiency in mesh expansion and non-manifold vertex detection, which currently requires scanning all boundary-edges \(\set E_{\text{boundary}}\); one solution is to index boundary-edges using an auxiliary octree activated only in affected regions;  
(2) developing methods to check if the mesh is a single edge-connected component without a full scan;  
(3) mitigating reconstruction errors from noisy point clouds, potentially using line-of-sight constraints to remove inaccurate facets.

\section{Acknowledgement}
This work was supported by the Research Council of Norway (RCN) through the Autonomous Robots for Ocean Sustainability (AROS) project (project number 304667) and the Center of Excellence, NTNU AMOS - Autonomous Marine Operations and Systems  (project number 223254), as well as the NTNU VISTA Centre for Autonomous Robotic Operations Subsea (CAROS).

We would like to thank NTNU AUR-lab, NTNU AMOS, NTNU VISTA CAROS for providing the multibeam echosounder (MBES) data.  A special note of appreciation goes to Dr. Ture Fronczek-Munter from Eelume AS for his assistance in obtaining the MBES data.

\appendix
\section{Volumetric vs Surface representation analysis}\label{app:volVsSurf}
In this appendix, we demonstrate why most volumetric next-best-view methods typically follow a \emph{define-first-detect-later} strategy, where candidate poses are proposed before detecting missing information. In contrast, our method adopts a \emph{detect-first-define-later} strategy.
To explain this distinction, we introduce the concept of a manifold.

For a rigorous definition of a smooth \(k\)-manifold in \(\mathbb{R}^n\), denoted as, \(\mani{k}{n}\), we refer the reader to \cite[Chapter 17.3, Definition 7]{adams2018calculus}. Also the term of boundary in this appendix is defined by \cite[Chapter 17.4, Page 986]{adams2018calculus}. However, rather than diving into formal definitions, we provide intuitive examples to help build a geometric understanding:
\begin{itemize}
    \item A 3-manifold in \(\mathbb{R}^3\), \(\mani{3}{3}\), is the entire \(\mathbb{R}^3\) space itself. It has no boundary.
    
    \item A 2-manifold in \(\mathbb{R}^3\), \(\mani{2}{3}\), is a continuous surface that either extends to infinity or forms a watertight (closed) surface. A hollow sphere is an example of a 2-manifold in \(\mathbb{R}^3\) without boundary. Such surfaces are continuous and differentiable everywhere.
    
    \item A 1-manifold in \(\mathbb{R}^3\), \(\mani{1}{3}\), is a continuous curve that either extends to infinity or forms a closed loop. A circle is an example of a 1-manifold in \(\mathbb{R}^3\). These curves are continuous and differentiable along their entire length and have no boundary.
\end{itemize}
However, in most practical robotic mapping applications, we cannot map the entire 3D space of our universe \(\mani{3}{3}\) using a volumetric representation. Nor can we assume that the surfaces we reconstruct are perfectly watertight, i.e., \(\mani{2}{3}\).
Therefore, a more practical focus is on subsets of these manifolds, denoted as \(\submani{k}{n} \subset \mani{k}{n}\), which are more relevant for real-world applications.
Again, rather than providing a formal definition \cite[Chapter 17.4, Definition 13]{adams2018calculus}, we present a few examples to offer an intuitive understanding of this concept.

\begin{itemize}
    \item A \(\submani{3}{3}\) can be a solid sphere, including all interior points and its boundary. The boundary of the solid sphere is a \(\mani{2}{3}\), namely the spherical surface.
    \item A \(\submani{2}{3}\) can be a hemispherical surface, including all points on the surface and its boundary. The boundary of the hemisphere is a \(\mani{1}{3}\), which is a circle.
    \item A \(\submani{1}{3}\) can be a curved line segment, including all points along the segment. The boundary of the segment is a \(\mani{0}{3}\), consisting of two points (the start and end).
\end{itemize}
From the above examples, we observe that the boundary of a \(\submani{k}{n}\) is a \(\mani{k-1}{n}\) and has no boundary. However, the boundary of a \(\submani{k}{n}\) can be more than a single  \(\mani{k-1}{n}\). We denote \(\psubmani{k}{n}\) as the set of boundaries of \(\mani{k}{n}\).

Now, we return to our robotic mapping application. In volumetric representations, the 3D space is typically partitioned into occupied, free, and unknown regions.
Assume we have a method capable of detecting unknown space in real-time. The main challenge lies in determining whether an unknown volume is observable. Let \(\submani{3}{3}\) denote all the occupied space, and let \(\psubmani{3}{3}\) represent the set of surfaces (boundaries) of \(\submani{3}{3}\), which may consist of one or more \(\mani{2}{3}\).

Since each \(\mani{2}{3}\) is an orientable, watertight surface and has no boundary, it has inside and outside (The well-known Jordan–Brouwer separation theorem). Therefore, we require an algorithm to determine whether the unknown volume lies inside any \(\mani{2}{3}\), in order to assess its observability. One such algorithm is the Even-Odd rule. However, this approach remains computationally expensive for real-time applications for a growing map, especially considering that a \(\mani{2}{3}\) can take on various complex shapes.

As a result, \emph{detect-first-define-later} strategy is uncommon in volumetric representations. Instead, \emph{define-first-detect-later} strategy are more prevalent. In this method, it is straightforward to first define a candidate sensor pose and then cast rays from the sensor to measure how much unknown volume can be observed. However, this typically requires random pose generation, like \cite{palomeras2019autonomous, bircher2016receding}.

In contrast, our method is surface-based and built upon \(\mani{2}{3}\) representations—more specifically, \(\submani{2}{3}\). We treat \(\psubmani{2}{3}\) as the missing information (holes) since \(\psubmani{2}{3}\) is a set of \(\mani{1}{3}\) and \(\mani{1}{3}\) is guaranteed to be closed (no boundary), which can be efficiently extracted using BPA. Given this, the work from \cite{yip2024active} can potentially be applied to determine suitable sensor poses for acquiring the missing information. Therefore, our approach follows a \emph{detect-first-define-later} strategy. Random pose generation is not needed.

We are aware that \(\mani{k}{n}\) and \(\submani{k}{n}\) are defined as smooth manifolds according to \cite{adams2018calculus}. In real-world applications, however, both volumetric methods and surface-based methods (such as ours) rely on discretized representations. Nevertheless, the smoothness property in the formal definition is not essential for the comparison presented in our analysis.

 \bibliographystyle{elsarticle-num-names.bst} 
 \bibliography{references.bib}

@article{niedzwiedzki2023idtmm,
  title = {{{IDTMM}}: {{Incremental Direct Triangle Mesh Mapping}}},
  shorttitle = {{{IDTMM}}},
  author = {Niedzwiedzki, Jakub and Lipinski, Piotr and Podsedkowski, Leszek},
  year = {2023},
  journal = {IEEE Robotics and Automation Letters},
  pages = {1--8},
  issn = {2377-3766, 2377-3774},
  doi = {10.1109/LRA.2023.3293751},
  urldate = {2023-07-14},
  langid = {english}
}

@inproceedings{kazhdan2006poisson,
  title={Poisson surface reconstruction},
  author={Kazhdan, Michael and Bolitho, Matthew and Hoppe, Hugues},
  booktitle={Proceedings of the fourth Eurographics symposium on Geometry processing},
  volume={7},
  number={4},
  year={2006}
}

@inproceedings{sheinin2016next,
  title={The next best underwater view},
  author={Sheinin, Mark and Schechner, Yoav Y},
  booktitle={Proceedings of the IEEE conference on computer vision and pattern recognition},
  pages={3764--3773},
  year={2016}
}

@article{mildenhall2021nerf,
  title={NeRF: Representing scenes as neural radiance fields for view synthesis},
  author={Mildenhall, Ben and Srinivasan, Pratul P and Tancik, Matthew and Barron, Jonathan T and Ramamoorthi, Ravi and Ng, Ren},
  journal={Communications of the ACM},
  volume={65},
  number={1},
  pages={99--106},
  year={2021},
  publisher={ACM New York, NY, USA}
}

@inproceedings{ruan2023slamesh,
  title={Slamesh: Real-time lidar simultaneous localization and meshing},
  author={Ruan, Jianyuan and Li, Bo and Wang, Yibo and Sun, Yuxiang},
  booktitle={2023 IEEE International Conference on Robotics and Automation (ICRA)},
  pages={3546--3552},
  year={2023},
  organization={IEEE}
}

@article{piazza2018real,
  title={Real-time cpu-based large-scale three-dimensional mesh reconstruction},
  author={Piazza, Enrico and Romanoni, Andrea and Matteucci, Matteo},
  journal={IEEE Robotics and Automation Letters},
  volume={3},
  number={3},
  pages={1584--1591},
  year={2018},
  publisher={IEEE}
}

@inproceedings{romanoni2015efficient,
  title={Efficient moving point handling for incremental 3d manifold reconstruction},
  author={Romanoni, Andrea and Matteucci, Matteo},
  booktitle={International Conference on Image Analysis and Processing},
  pages={489--499},
  year={2015},
  organization={Springer}
}

@INPROCEEDINGS{newcombe2011kinectfusion,
  author={Newcombe, Richard A. and Izadi, Shahram and Hilliges, Otmar and Molyneaux, David and Kim, David and Davison, Andrew J. and Kohi, Pushmeet and Shotton, Jamie and Hodges, Steve and Fitzgibbon, Andrew},
  booktitle={2011 10th IEEE International Symposium on Mixed and Augmented Reality}, 
  title={KinectFusion: Real-time dense surface mapping and tracking}, 
  year={2011},
  volume={},
  number={},
  pages={127-136},
  doi={10.1109/ISMAR.2011.6092378}}

@inproceedings{curless1996volumetric,
  title={A volumetric method for building complex models from range images},
  author={Curless, Brian and Levoy, Marc},
  booktitle={Proceedings of the 23rd annual conference on Computer graphics and interactive techniques},
  pages={303--312},
  year={1996}
}

@inproceedings{vizzo2021poisson,
  title={Poisson surface reconstruction for LiDAR odometry and mapping},
  author={Vizzo, Ignacio and Chen, Xieyuanli and Chebrolu, Nived and Behley, Jens and Stachniss, Cyrill},
  booktitle={2021 IEEE international conference on robotics and automation (ICRA)},
  pages={5624--5630},
  year={2021},
  organization={IEEE}
}

@article{schops2019surfelmeshing,
  title={Surfelmeshing: Online surfel-based mesh reconstruction},
  author={Sch{\"o}ps, Thomas and Sattler, Torsten and Pollefeys, Marc},
  journal={IEEE transactions on pattern analysis and machine intelligence},
  volume={42},
  number={10},
  pages={2494--2507},
  year={2019},
  publisher={IEEE}
}

@inproceedings{oleynikova2017voxbloxa,
  title = {Voxblox: {{Incremental 3D Euclidean Signed Distance Fields}} for on-Board {{MAV}} Planning},
  shorttitle = {Voxblox},
  booktitle = {2017 {{IEEE}}/{{RSJ International Conference}} on {{Intelligent Robots}} and {{Systems}} ({{IROS}})},
  author = {Oleynikova, Helen and Taylor, Zachary and Fehr, Marius and Siegwart, Roland and Nieto, Juan},
  year = {2017},
  month = sep,
  pages = {1366--1373},
  publisher = {IEEE},
  address = {Vancouver, BC},
  doi = {10.1109/IROS.2017.8202315},
  urldate = {2024-07-05},
  isbn = {978-1-5386-2682-5},
  langid = {english}
}

@inproceedings{yu2019incremental,
  title={Incremental Poisson Surface Reconstruction for large scale three-dimensional modeling},
  author={Yu, Qiang and Sui, Wei and Wang, Ying and Xiang, Shiming and Pan, Chunhong},
  booktitle={Pattern Recognition and Computer Vision: Second Chinese Conference, PRCV 2019, Xi’an, China, November 8--11, 2019, Proceedings, Part III 2},
  pages={442--453},
  year={2019},
  organization={Springer}
}

@article{palomeras2019autonomous,
  title={Autonomous exploration of complex underwater environments using a probabilistic next-best-view planner},
  author={Palomeras, Narc{\'\i}s and Hurt{\'o}s, Natalia and Vidal, Eduard and Carreras, Marc},
  journal={IEEE Robotics and Automation Letters},
  volume={4},
  number={2},
  pages={1619--1625},
  year={2019},
  publisher={IEEE}
}

@inproceedings{bircher2016receding,
  title={Receding horizon ``next-best-view" planner for 3d exploration},
  author={Bircher, Andreas and Kamel, Mina and Alexis, Kostas and Oleynikova, Helen and Siegwart, Roland},
  booktitle={2016 IEEE international conference on robotics and automation (ICRA)},
  pages={1462--1468},
  year={2016},
  organization={IEEE}
}

@article{yip2024active,
  title={Active Pose-Autonomous Hole-Filling Techniques for 3D Surface Reconstruction in Underwater Exploration},
  author={Yip, Mauhing and Schellewald, Christian and Gambin, Timmy and Stahl, Annette},
  journal={IFAC-PapersOnLine},
  volume={58},
  number={20},
  pages={293--300},
  year={2024},
  publisher={Elsevier}
}

@article{yip2024robust,
  title={Robust hole-detection in triangular meshes irrespective of the presence of singular vertices},
  author={Yip, Mauhing and Stahl, Annette and Schellewald, Christian},
  journal={Computer-Aided Design},
  volume={170},
  pages={103696},
  year={2024},
  publisher={Elsevier}
}

@article{digne2014analysis,
  title={An analysis and implementation of a parallel ball pivoting algorithm},
  author={Digne, Julie},
  journal={Image Processing On Line},
  volume={4},
  pages={149--168},
  year={2014}
}

@article{bernardini2002ball,
  title={The ball-pivoting algorithm for surface reconstruction},
  author={Bernardini, Fausto and Mittleman, Joshua and Rushmeier, Holly and Silva, Cl{\'a}udio and Taubin, Gabriel},
  journal={IEEE transactions on visualization and computer graphics},
  volume={5},
  number={4},
  pages={349--359},
  year={2002},
  publisher={IEEE}
}

@InProceedings{liljeback2017eelume,
  author       = {Liljeb{\"a}ck, P{\aa}l and Mills, Richard},
  booktitle    = {Oceans 2017-Aberdeen},
  title        = {Eelume: A flexible and subsea resident IMR vehicle},
  year         = {2017},
  organization = {IEEE},
  pages        = {1--4},
}

@book{grimaldi2006discrete,
  title={Discrete and Combinatorial Mathematics, 5/e},
  author={Grimaldi, Ralph P},
  year={2006},
  publisher={Pearson Education India}
}

@article{campos2015surface,
  title={A surface reconstruction method for in-detail underwater 3D optical mapping},
  author={Campos, Ricard and Garcia, Rafael and Alliez, Pierre and Yvinec, Mariette},
  journal={The International Journal of Robotics Research},
  volume={34},
  number={1},
  pages={64--89},
  year={2015},
  publisher={SAGE Publications Sage UK: London, England}
}

@book{adams2018calculus,
  title={Calculus: a complete course, 9th edition},
  author={Adams, Robert A and Essex, Christopher},
  year={2018},
  publisher={Pearson}
}

@article{hornung2013octomap,
  title={OctoMap: An efficient probabilistic 3D mapping framework based on octrees},
  author={Hornung, Armin and Wurm, Kai M and Bennewitz, Maren and Stachniss, Cyrill and Burgard, Wolfram},
  journal={Autonomous robots},
  volume={34},
  pages={189--206},
  year={2013},
  publisher={Springer}
}

@INPROCEEDINGS{ReAqROVIO,
  author={Singh, Mohit and Alexis, Kostas},
  booktitle={2024 IEEE/RSJ International Conference on Intelligent Robots and Systems (IROS)}, 
  title={Online Refractive Camera Model Calibration in Visual Inertial Odometry}, 
  year={2024},
  volume={},
  number={},
  pages={12609-12616},
  doi={10.1109/IROS58592.2024.10802302}
}

@article{kerbl20233d,
  title={3D Gaussian splatting for real-time radiance field rendering.},
  author={Kerbl, Bernhard and Kopanas, Georgios and Leimk{\"u}hler, Thomas and Drettakis, George},
  journal={ACM Trans. Graph.},
  volume={42},
  number={4},
  pages={139--1},
  year={2023}
}

@article{zhou2018open3d,
  title={Open3D: A modern library for 3D data processing},
  author={Zhou, Qian-Yi and Park, Jaesik and Koltun, Vladlen},
  journal={arXiv preprint arXiv:1801.09847},
  year={2018}
}

@article{zeeman1966introduction,
  title={An introduction to topology: The classification theorem for surfaces},
  author={Zeeman, Eric-Christopher},
  journal={(No Title)},
  year={1966}
}

@inproceedings{sequin2012moebius,
  title={From Moebius Bands to Klein-Knottles},
  author={S{\'e}quin, Carlo H},
  booktitle={Proceedings of Bridges 2012: Mathematics, Music, Art, Architecture, Culture},
  pages={93--102},
  year={2012}
}

@INPROCEEDINGS{DeepVL,
  author={Singh, Mohit and Alexis, Kostas},
  booktitle={2025 IEEE International Conference on Robotics and Automation (ICRA)}, 
  title={DeepVL: Dynamics and Inertial Measurements-based Deep Velocity Learning for Underwater Odometry}, 
  year={2025},
  volume={},
  number={},
  pages={1-7},
  doi={10.1109/ICRA55743.2025.11128041}}
 %\bibliography{cas-refs}





% \end{thebibliography}
\end{document}